\documentclass[12pt]{article}
\usepackage{epsf}
\usepackage{latexsym}
\usepackage{cite}
\newcommand{\beps}{\mbox{\boldmath$\varepsilon$}}
\newcommand{\bk}{\mbox{$\mathbf k$}} %%%{\mbox{\boldmath$k$}}
\newcommand{\bq}{\mbox{$\mathbf q$}} %%%{\mbox{\boldmath$q$}}

\begin{document}
%%%%%%%%%%%%%%%%%%%%%%%%%

\begin{flushright}
{\large PSI-PR-00-09} \\
May 17, 2000
\\[15mm]

\end{flushright}

%%%%%%%%%%%%%%%%%%%%%%%%%%%%%%%%%%%%%%%%%%%%%%%%%%%%%%%%%%%%%%%%%%%%%%%%%%%
\begin{center}
{\Large\sffamily\bfseries
  The radiative decay $\phi\to\gamma\pi\pi$ in a coupled channel model
  and the structure of $f_0(980)$
}\\[5mm]
{\sc V.E.~Markushin}\\[5mm]
{\it Paul Scherrer Institute, 5232 Villigen PSI, Switzerland}\\[5mm]
\end{center}

%%%%%%%%%%%%%%%%%%%%%%%%%%%%%%%%%%%%%%%%%%%%%%%%%%%%%%%%%%%%%%%%%%%%%%%%%%%
\begin{abstract}
A coupled channel model is used to study the nature of the scalar
mesons produced in the decay $\phi\to\gamma\pi\pi$.
The $K\bar{K}$ molecular picture of $f_0(980)$ is found to be in a
good agreement with the recent experimental data from SND and
CMD-2.  The structure of the light scalar mesons is elucidated by
investigating the $S$-matrix poles and the $q\bar{q}$
spectral density.
\end{abstract}

%%%%%%%%%%%%%%%%%%%%%%%%%%%%%%%%%%%%%%%%%%%%%%%%%%%%%%%%%%%%%%%%%%%%%%%%%%%
\section{Introduction}
\label{INT}
%%%%%%%%%%%%%%%%%%%%%%%%%%%%%%%%%%%%%%%%%%%%%%%%%%%%%%%%%%%%%%%%%%%%%%%%%%%

  The production of scalar mesons in radiative decays is a valuable source
of information on hadron spectroscopy.
It was argued \cite{ADS80,AI89} that the branching ratio for
$\phi\to\gamma f_0(980)$ can be used to
make a unique choice among different models of $f_0(980)$:
a conventional quark-antiquark state,
an exotic $qq\bar{q}\bar{q}$ state \cite{Ja77},
and a $K\bar{K}$ molecule \cite{WI82}.
Such a possibility to resolve the long debated problem of the $f_0(980)$
structure using a single partial width looks very attractive.  However,
it was argued in \cite{CIK93} that the dependence of the theoretical
predictions for the $\phi\to\gamma f_0(980)$ decay on the
$f_0(980)$ structure is partly due to differences in modeling.
Recent measurements of the radiative decays $\phi\to\gamma \pi^0\pi^0$
by SND \cite{SND98}
and $\phi\to\gamma \pi^+\pi^-, \gamma\pi^0\pi^0$ by CMD-2 \cite{CMD99a,CMD99b}
in Novosibirsk have made it possible to confront alternative models of
the light scalar mesons with experimental data.

   The goal of this paper is to reanalyze the calculation of the
decay width for $\phi\to\gamma f_0(980)$ in coupled channel models
where the $f_0(980)$ state arises as a dynamical state
(a $K\bar{K}$ molecule which may also have a substantial admixture of
a quark-antiquark component).
   Since the $\phi$ meson is nearly a pure $s\bar{s}$ state, the
decay $\phi\to\gamma \pi\pi$ is an OZI--rule violating process which
is expected to proceed via a two--step mechanism with intermediate
$K\bar{K}$ states.  Therefore this decay is well suited for probing
the $K\bar{K}$ content of the scalar mesons.
  The $K\bar{K}$ molecular state was originally proposed in the
potential quark model \cite{WI82,GI85,WI90}.
  A dynamical state close to $K\bar{K}$ threshold is also introduced
in the coupled channel models \cite{CDL89,KLM94,LMZ98,KLL99}
and in the meson exchange interaction models \cite{Lo90,JPHS95}.
  A state strongly coupled to the $s\bar{s}$ and
$K\bar{K}$  channels near the $K\bar{K}$ threshold was as well
found in the unitarized quark model \cite{To95,To96}.
  The coupled-channel model derived from the lowest order chiral Lagrangian
\cite{OO97,OO99} produces a scalar state dominated by the $K\bar{K}$ channel.
  General discussions of the nature of the $f_0(980)$ can be found in
\cite{WI90,LMZ98,To95,RPP98,MP93,Ma99} and references therein.
% Pa95 ,Frascati ?

  Our approach is based on a coupled channel model (CCM) for the $\pi\pi$ and
$K\bar{K}$ systems which is similar to the one studied in
\cite{LMZ98}. The calculation of the decay $\phi\to\gamma
f_0(980)$ for point-like particles is summarized in Sec.~\ref{DA}.
  The details of the coupled channel model are given in Sec.~\ref{CCM},
and the model parameters are determined from a fit to the $\pi\pi$ scattering data.
The reaction $\phi\to\gamma\pi\pi$ in a CCM framework is studied in Sec.~\ref{DACCM}.
  The analytic structure of the $\pi\pi$ and $K\bar{K}$ scattering amplitudes
is investigated in Sec.~\ref{POLES}.
  The mixing between the two--meson and $q\bar{q}$ channels is discussed
in Sec.~\ref{Mixingqqbar}.
  In Sec.~\ref{Disc} the physical properties of the scalar mesons in the
model proposed are discussed and compared to other approaches in
the literature.
The details of the formalism are collected in the Appendices.

%%%%%%%%%%%%%%%%%%%%%%%%%%%%%%%%%%%%%%%%%%%%%%%%%%%%%%%%%%%%%%%%%%%%%%%%%%%
\section{The Decay $\phi\to\gamma f_0$}
\label{DA}
%%%%%%%%%%%%%%%%%%%%%%%%%%%%%%%%%%%%%%%%%%%%%%%%%%%%%%%%%%%%%%%%%%%%%%%%%%%

For the benefit of the reader, we begin with a brief summary of the
formulas describing radiative transitions between vector and scalar
states.
  The amplitude of the radiative $\phi$ decay into the scalar meson $f_0$
has the following structure which is imposed by gauge invariance:
\begin{eqnarray}
 M(\phi\to\gamma f_0) & = &
 \epsilon_{\phi}^{\mu} \epsilon_{\gamma}^{\nu}
 (p_{\nu} q_{\mu} - g_{\nu\mu} (p\cdot q)) H(p^2,(p-q)^2)
 \label{MphigS}
\end{eqnarray}
where ($\epsilon_{\phi}$, $p$) and ($\epsilon_{\gamma}$, $q$) are
the polarizations and four-momenta of the $\phi$ and $\gamma$,
correspondingly, and $H(p^2,(p-q)^2)$ is the scalar invariant
amplitude which depends on the invariant masses of the initial and
final mesons. The polarization vectors satisfy the constraints
$\epsilon_{\phi}\cdot p=0$ and $\epsilon_{\gamma}\cdot q =0$.
Using the three-dimensional gauge
$\epsilon_{\gamma}=(0,\beps_{\gamma})$ in
the center-of-mass system (CMS) one gets the amplitude
\begin{eqnarray}
 M(\phi\to\gamma f_0) & = &
 ( \beps_{\phi} \beps_{\gamma} )
 m_{\phi} \omega H(p^2, (p-q)^2)
 \label{MphigScms}
\end{eqnarray}
where $m_{\phi}$ is the $\phi$ mass, $\omega$ is the photon energy
in the CMS, and $\beps_{\phi}$ and $\beps_{\gamma}$ are the $\phi$
and $\gamma$ three--dimensional polarization vectors in the CMS, respectively.

%%%%%%%%%%%%%%%%%%%%%%%%%%%%%%%%%%%%%%%%%%%%%%%
\begin{figure}[ht]
\begin{center}
\mbox{(a) \hspace{80mm} (b)}\\[-5mm]
\mbox{
\mbox{\epsfxsize=50mm \epsffile{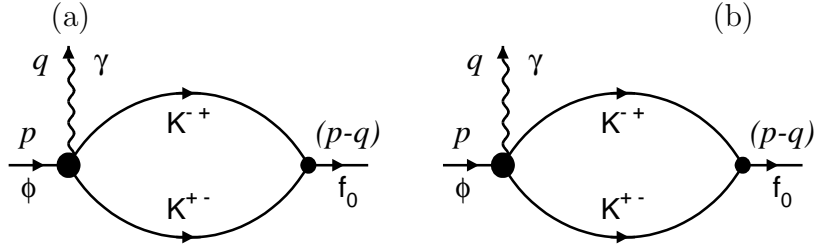}} \hspace{5mm}
\mbox{\epsfxsize=50mm \epsffile{figbphigammaf0.eps}}
}
\caption{\label{FigMphigS}%
The minimal gauge-invariant set of diagrams for the decay
$M(\phi\to\gamma f_0)$ in the case of point-like particles:
(a) the loop radiation; (b) the contact term.
}
\end{center}
\end{figure}
%%%%%%%%%%%%%%%%%%%%%%%%%%%%%%%%%%%%%%%%%%%%%%%

  For point-like particles, the amplitude $M(\phi\to\gamma f_0)$ is given by
the diagrams in Fig.\ref{FigMphigS}
(see \cite{CIK93} and references therein).
While the diagram Fig.\ref{FigMphigS}(a) corresponding to loop radiation is
overall logarithmically divergent, its contribution to the $p_{\nu} q_{\mu}$ term
in Eq.(\ref{MphigS}) is finite.  Gauge invariance enforces the
appearance of the seagull diagram Fig.\ref{FigMphigS}(b)
which contributes only to the $g_{\nu\mu} (p\cdot q)$ term in Eq.(\ref{MphigS}).
Since the sum of the loop--radiation and seagull terms is gauge invariant,
one obtains the total amplitude $M(\phi\to\gamma S)$ by calculating
only the $p_{\nu} q_{\mu}$ term of the loop radiation.
  The result for the scalar invariant amplitude is \cite{LP90}
\begin{eqnarray}
  H(p^2,(p-q)^2) & = &
  \frac{e g_{\phi} g_{f_0KK}}{2\pi^2 m_K^2} I(a,b)
\label{HphigS}
\\
  a = \frac{m_{\phi}^2}{m_K^2}
  & , &
  b = \frac{m_{f_0}^2}{m_K^2}
\end{eqnarray}
where $g_{\phi}$ and $g_{f_0KK}$ are the $\phi K^+K^-$ and $f_0 K^+K^-$ coupling
constants and the function $I(a,b)$ is defined in \ref{AppIab}.
%%% Appendix~\ref{AppIab}
The radiative decay width is
\begin{eqnarray}
 \Gamma(\phi\to\gamma f_0) & = &
 \frac{\omega^3 |H(p^2,(p-q)^2)|}{12\pi} =
%   \frac{\alpha g_{\phi}^2 g_{f_0KK}^2}{6 (2\pi)^4}  % approx.
%   \frac{(m_{\phi}-m_{f_0})^3}{ m_{\phi}^3 m_K^4}
%   |I(a,b)|^2  =
\\
& = &
   \frac{\alpha g_{\phi}^2 g_{f_0KK}^2}{3 (2\pi)^4}
   \frac{\omega}{m_{\phi}^2}
   |(a-b)I(a,b)|^2 \quad .
\label{GphigS}
\end{eqnarray}

The generalization of Eqs.~(\ref{MphigS},\ref{HphigS},\ref{GphigS})
to the case $\phi\to\gamma\pi\pi$ where the $\pi\pi$ system has total
angular momentum $J=0$ and isospin $I=0$ is straightforward \cite{MHOT99}:
\begin{eqnarray}
 M(\phi\to\gamma \pi\pi) & = &
 \epsilon_{\phi}^{\mu} \epsilon_{\gamma}^{\nu}
 (p_{\nu} q_{\mu} - g_{\nu\mu} (p\cdot q)) H_{\pi\pi}(p^2,(p-q)^2) \quad . 
 \label{Mphigpipi}
\end{eqnarray}
Here the scalar invariant amplitude $H_{\pi\pi}(p^2,(p-q)^2)$ is
given by (compare with Eq.(\ref{HphigS}))
\begin{eqnarray}
 H_{\pi\pi}(p^2,(p-q)^2) & = &
 \frac{e g_{\phi}}{2\pi^2 m_K^2} I(a,b) \; t_{K^+K^--\pi\pi}
\label{Hphigpipi}
\end{eqnarray}
and $t_{K^+K^--\pi\pi}$ is the $J=0$ part of the $T$-matrix for
the $K^+K^-\to\pi\pi$ scattering.
  The $\pi\pi$ invariant mass distribution has the form
%% done 09/05/2000
\begin{eqnarray}
   \frac{d\Gamma}{d M_{\pi^+\pi^-}} = 2 \frac{d\Gamma}{d M_{\pi^0\pi^0}}
   & = &
   \frac{\alpha g_{\phi}^2 \; \omega }{18 (2\pi)^6 m_{\phi}^2}
   |(a-b)I(a,b)|^2 |t^0_{K\bar{K}-\pi\pi}|^2 k_{\pi\pi}
\label{dGphigpipi}
\end{eqnarray}
where
$t^0_{K\bar{K}-\pi\pi}=\sqrt{6}\;t_{K^+K^--\pi^+\pi^-}$ is the isoscalar
$K\bar{K}\to\pi\pi$ amplitude and
$k_{\pi\pi}$ is the relative momentum of the pions in the final state:
\begin{eqnarray}
   k_{\pi\pi} & = & \sqrt{M_{\pi\pi}^2/4 - m_{\pi}^2}
  \quad , \quad
  M_{\pi\pi}^2 = (p-q)^2 \quad .
\end{eqnarray}

Equation (\ref{dGphigpipi}) leads to Eq.(\ref{GphigS}) in
the Breit--Wigner (BW) approximation for the $K^+K^-\to\pi\pi$
scattering amplitude
\begin{eqnarray}
  t_{K^+K^--\pi\pi} & = &
  \frac{g_{f_0KK} g_{f_0\pi\pi}}{M_{\pi\pi}^2 - (M_{f_0} - i\Gamma_{f_0}/2)^2}
\end{eqnarray}
under the assumption that the integral over the $\pi\pi$ mass spectrum
is saturated by the narrow resonance
with mass $M_{f_0}$ and width $\Gamma_{f_0}$
(see \ref{AppIab}).

According to Eq.(\ref{Hphigpipi}), the hadronic part of the amplitude,
which contains the information about the scalar mesons, is factored out
in the form of the $T$-matrix for the $K^+K^-\to\pi\pi$ scattering,
with both the physical region ($M_{\pi\pi} \geq 2m_K$) and the unphysical region
($M_{\pi\pi} < 2m_K$) being relevant to the $\phi\to\gamma\pi\pi$ decay.
It is known from the studies of scalar mesons, see \cite{LMZ98,MP93,Ma99,AMP87}
and references therein, that the analytical structure of the scalar--isoscalar
amplitudes near the $K\bar{K}$ threshold is far from being a trivial BW resonance.
Therefore a coupled channel model of the $K^+K^-\to\pi\pi$ scattering
is required to describe the decay $\phi\to\gamma\pi\pi$ beyond the BW
approximation.

%%%%%%%%%%%%%%%%%%%%%%%%%%%%%%%%%%%%%%%%%%%%%%%%%%%%%%%%%%%%%%%%%%%%%%%%%%%
\section{The $\pi\pi-K\bar{K}$ Coupled Channel Model}
\label{CCM}
%%%%%%%%%%%%%%%%%%%%%%%%%%%%%%%%%%%%%%%%%%%%%%%%%%%%%%%%%%%%%%%%%%%%%%%%%%%

To describe the interaction in the $\pi\pi-K\bar{K}$ system
with total angular momentum $J=0^{++}$ and isospin $I^G=0^+$
we exploit a coupled channel model similar to that of \cite{LMZ98}.
The two scattering channels, 1 and 2, correspond to the $\pi\pi$ and $K\bar{K}$
systems and channel 3 contains a single $q\bar{q}$ bound state.
The $T$-matrix, as a function of the invariant mass squared $s$, is
defined by the Lippmann-Schwinger equation
\begin{eqnarray}
    \mbox{\boldmath$T$}(s) & = & \mbox{\boldmath$V$}
           + \mbox{\boldmath$V$} \mbox{\boldmath$G$}^0(s) \mbox{\boldmath$T$}
\label{T}
\end{eqnarray}
where $\mbox{\boldmath$G$}^0(s)$ is the free Green function.

The interaction potentials are taken in separable form:
\begin{eqnarray}
   \mbox{\boldmath$V$} & = &
   \left( \matrix{
     v_{11}(s) |1 \rangle\langle 1| & v_{12}(s) |1 \rangle\langle 2| &
                                  g_{13} |1 \rangle\langle q\bar{q}| \cr
     v_{12}(s) |2 \rangle\langle 1| & v_{22}(s) |2 \rangle\langle 2| &
                                  g_{23} |2 \rangle\langle q\bar{q}| \cr
     g_{13}|q\bar{q} \rangle\langle 1| &
     g_{23}|q\bar{q} \rangle\langle 2|  & 0 \cr }
   \right)
\label{Vsep}
\end{eqnarray}
where the form factors in channel 1 and 2
depend on the corresponding relative three--momentum $k$:
\begin{eqnarray}
   \langle k|1 \rangle & = & \xi_1(k) = \frac{\lambda_1^{2}}{k^2+\lambda_1^2} \\
   \langle k|2 \rangle & = & \xi_2(k) = \frac{\lambda_2^{2}}{k^2+\lambda_2^2}
\label{FF}
\end{eqnarray}
In \cite{LMZ98}, the potentials $v_{ij}(s)$ were assumed to be
energy independent, and the chiral symmetry constraints on the
scattering amplitude were strictly imposed only in the $\pi\pi$
channel by adjusting the strength of $v_{11}$ so that the Adler
zero is at the correct position. In the present case, it is
essential to ensure a correct behaviour of the $K\bar{K}-\pi\pi$
scattering amplitude not only in the physical scattering region
but also down to the $\pi\pi$ threshold.
We found it easier to impose the chiral symmetry constraints by
using energy dependent potentials\footnote{The unitarization of
the lowest order in chiral perturbation theory goes along a
similar way, see \protect\cite{OO97} and references therein.}. The
energy dependence is taken in the form
\begin{eqnarray}
    v_{11}(s) & = &  b-g_{11}s   %%% (s - m_{\pi}^2/2)
\label{v11}
\\
    v_{12}(s) & = &  -g_{12}s
\label{v12}
\\
    v_{22}(s) & = &  -g_{22}s  \quad  .
\label{v22}
\end{eqnarray}
With our choice of interaction (\ref{Vsep}-\ref{v22})
the analytical solution for the $T$-matrix can be easily obtained.
Further details of the model are given in \ref{AppCCM}.
%%% Appendix~\ref{AppCCM}
%%% \marginpar{\it CCM --- ChPT\\ parameters} ???
  As in \cite{LMZ98} we assume that the diagonal interaction in
the $K\bar{K}$ channel produces a weakly bound state in the absence of
coupling to the other channels, thus simulating a ``molecular'' origin of
the $f_0(980)$ resonance.
The state $|q\bar{q}\rangle$ in channel 3 has a bare mass $M_r > 2m_K$.

%%%%%%%%%%%%%%%%%%%%%%%%%%%%%
\begin{figure}[hbt]
\begin{center}
\mbox{(a) \hspace{70mm} (b)}\\
\mbox{
\mbox{\epsfxsize=70mm \epsffile{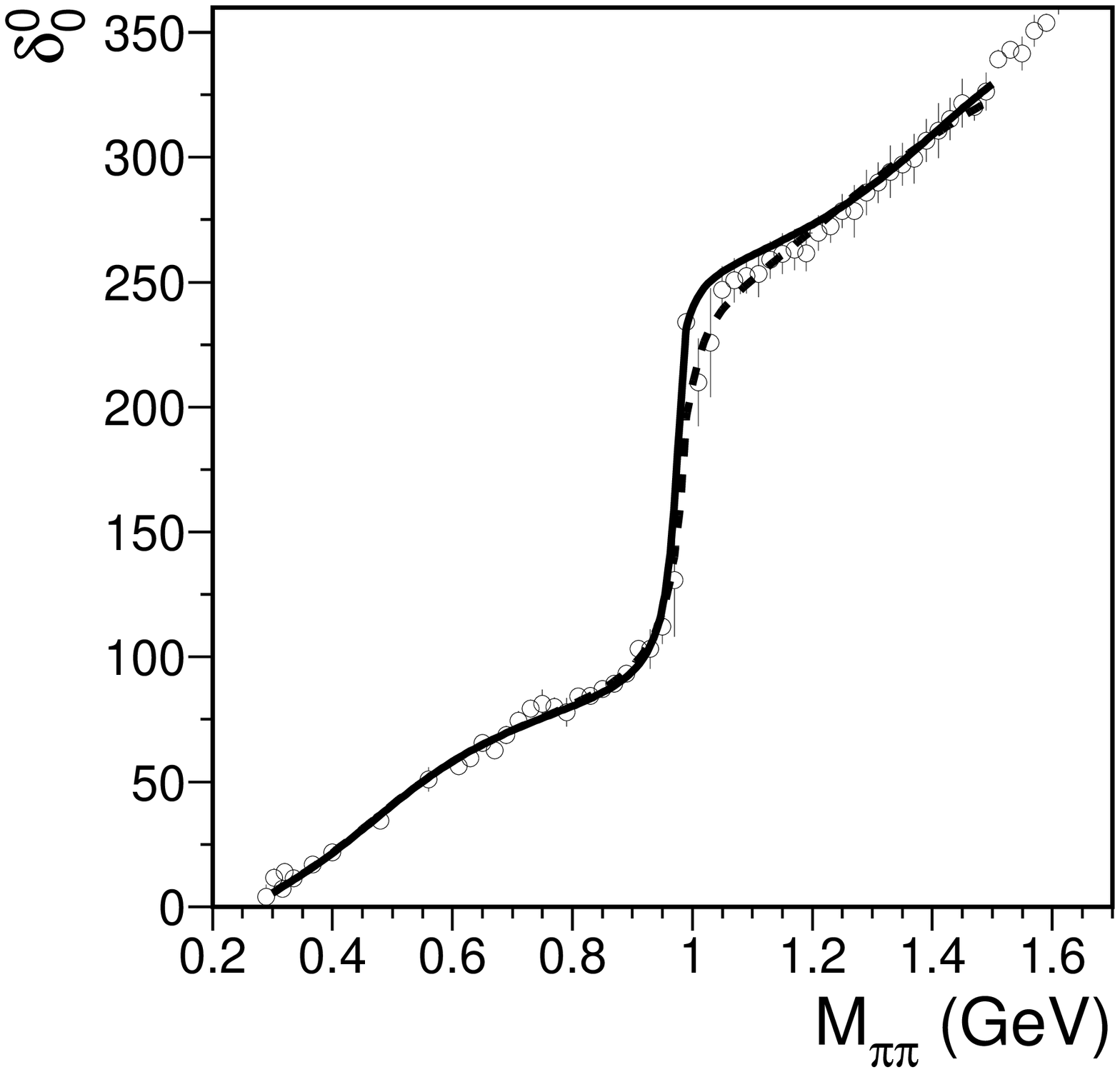}} \hspace{5mm}
\mbox{\epsfxsize=70mm \epsffile{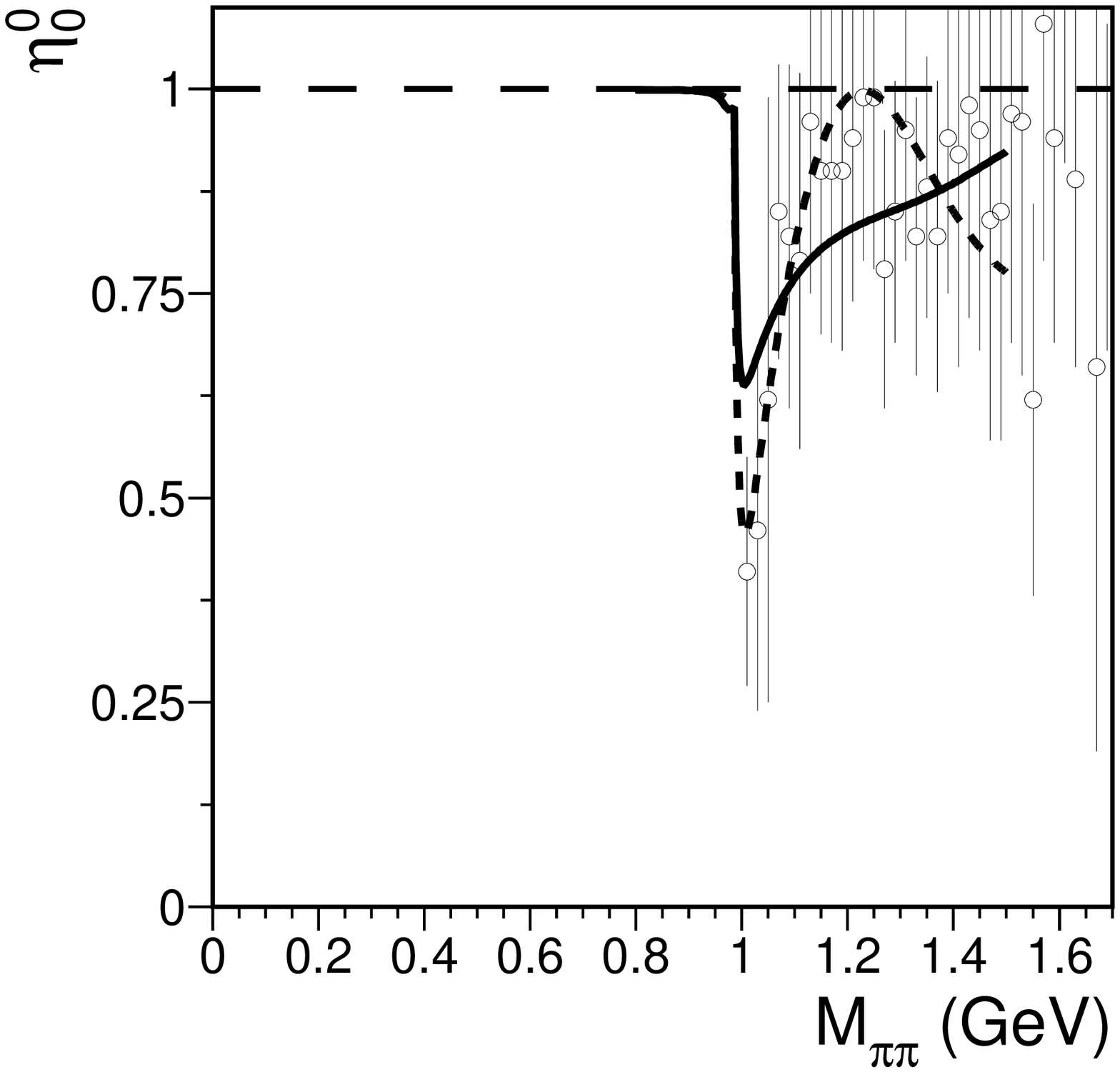}}
}
\end{center}
\caption{\label{Fit}%
The scattering phase $\delta_0^0$ and the inelasticity
parameter $\eta_0^0$ for the $S$-wave $\pi\pi$ scattering vs. 
the $\pi\pi$ invariant mass.
The experimental points are from \protect\cite{Gr74,Oc74,Ro77}.
The curves show the fits in our model:
fit 1 (the dashed lines) involves only the $\pi\pi$ scattering data,
fit 4 (the solid lines) includes in addition the data for 
$\phi\to\gamma\pi^0\pi^0$ \cite{SND98,CMD99b}.
}
\end{figure}
%%%%%%%%%%%%%%%%%%%%%%%%%%%%%

%%%%%%%%%%%%%%%%%%%%%%%%%%%%%
\begin{table}[htb]
\caption{\label{TFit}%
The model parameters obtained from the data fits.
The fit 1 is based on using only the $\pi\pi$ scattering data
(the phase $\delta_0^0$ and the inelasticity $\eta_0^0$).
The fits 2--4 include in addition the mass distribution $d\Gamma/dM_{\pi\pi}$
in the decay $\phi\to\gamma\pi^0\pi^0$:
data from SND \cite{SND98} in fit~2, from CMD-2 \cite{CMD99b} in fit~3,
and the combined data \cite{SND98,CMD99b} in fit~4.
The value of $\lambda_2$ was fixed in all fits.
}
\begin{center}
%%%%%%%%%%%%%%%%%%%%%%%%%%%%%
% Fit 1 = FitH312z4  May 09, 2000
% Fit 2 = FitH312y4  May 09, 2000
% Fit 3 = FitH312c   May 15, 2000
% Fit 4 = FitH312A   May 15, 2000 
% g_13 := 2*g13
% g_23 := 2*g23 
%%%%%%%%%%%%%%%%%%%%%%%%%%%%%
\begin{tabular}{|c|ccccccccc|} \hline
fit & $b$ & $g_{11}$  &  $g_{22}$ & $g_{12}$  & $\lambda_1$ & $\lambda_2$
    & $g_{13}$  & $g_{23}$  & $M_r$
\\
    & GeV$^{-1}$  & GeV$^{-3}$  & GeV$^{-3}$ & GeV$^{-3}$ & GeV     & GeV
    & GeV$^{1/2}$ & GeV$^{1/2}$ & GeV
\\ \hline
(1) & 5.251  & 3.564     &  5.326    &  3.741     &  0.477  &  0.7
    &  2.670    &  0.464    &  1.136
\\ \hline
(2) & 6.075  &  2.953    &  3.945    &  -0.121    &  0.499  &  0.7
    &  2.708    &  0.716    &  1.094
\\ \hline
(3) & 5.188  &  2.944    &  3.683    &  -0.270    &  0.538  &  0.7
    &  2.576    &  0.690    &  1.111
\\ \hline
(4) & 5.299  &  2.954    &  3.725    &  -0.341    &  0.529  &  0.7
    &  2.588    &  0.702    &  1.105
\\ \hline
%\\ \hline
\end{tabular}
\end{center}
\end{table}
%%%%%%%%%%%%%%%%%%%%%%%%%%%%%

%%%%%%%%%%%%%%%%%%%%%%%%%%%%%
\begin{figure}[hbt]
\begin{center}
\mbox{\epsfxsize=80mm \epsffile{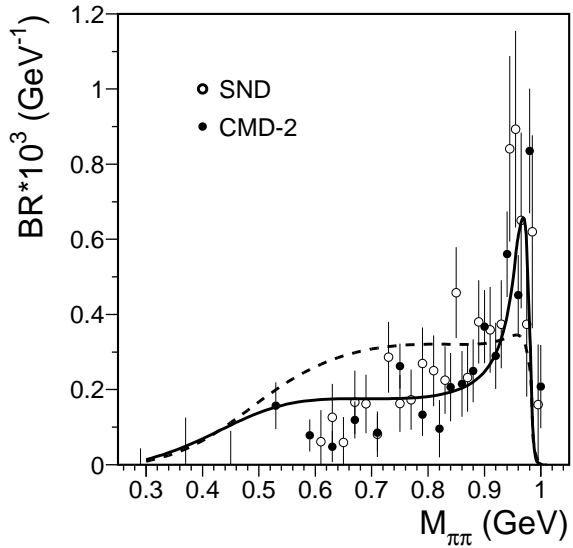}}
\end{center}
\caption{\label{FitGamma}%
The $\pi\pi$ invariant mass distribution in the
decay $\phi\to\gamma\pi^0\pi^0$.
The curves are the results from our model:
the dashed line corresponds to fit 1, the solid line to fit 4.
The experimental points are from \protect\cite{SND98} and \protect\cite{CMD99b}.
}
\end{figure}
%%%%%%%%%%%%%%%%%%%%%%%%%%%%%

%%%%%%%%%%%%%%%%%%%%%%%%%%%%%%%%%%%%%%%%%%%%%%%
\begin{figure}[htb]
\begin{center}
\mbox{\epsfxsize=70mm \epsffile{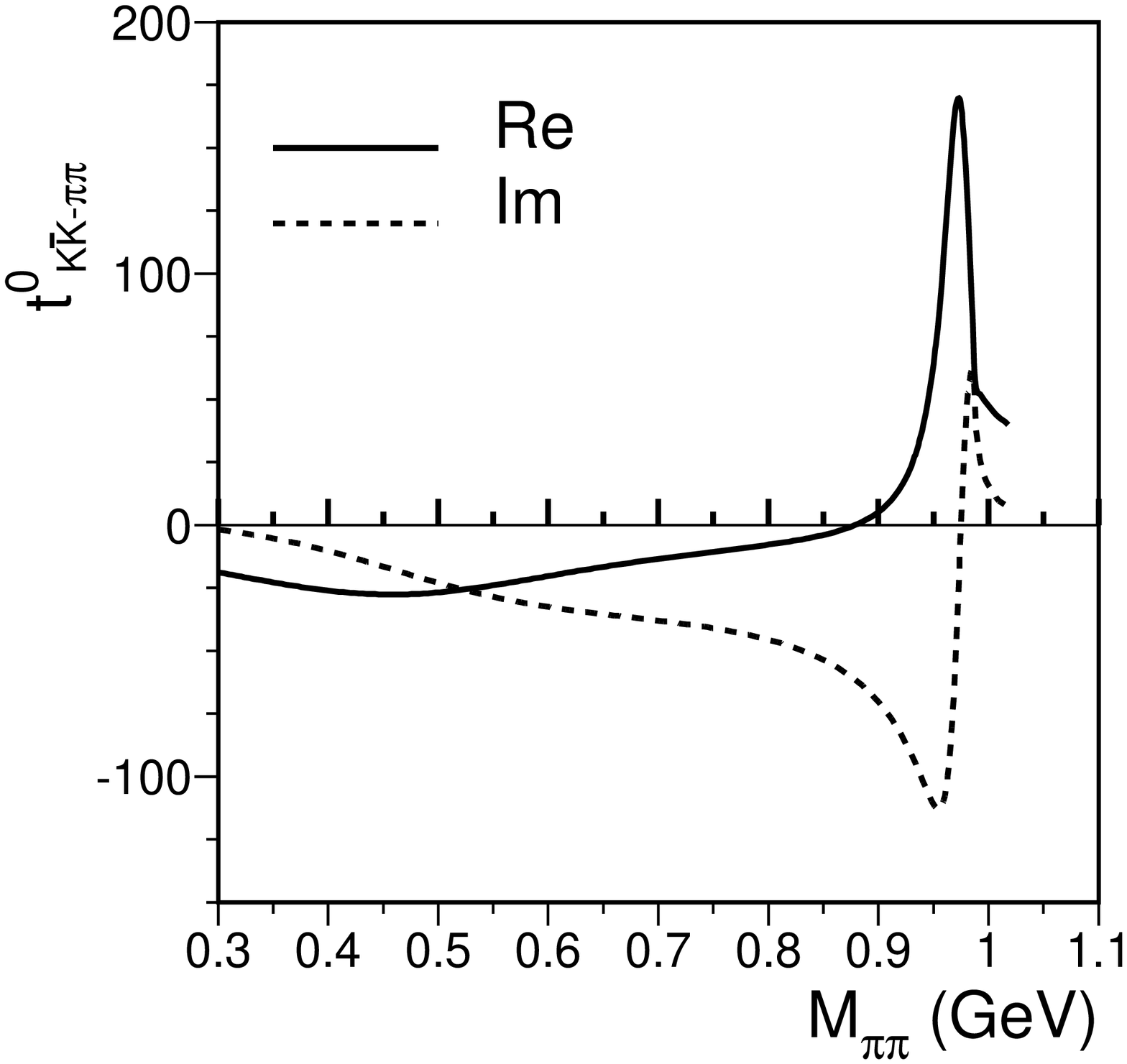}}
\end{center}
\caption{\label{FigT12}%
The $T$--matrix for the $K\bar{K}-\pi\pi$ scattering ($J=I=0$)
vs. the invariant mass of the $\pi\pi$ system $M_{\pi\pi}$.
The model parameters correspond to the fit 4 in Table~\protect\ref{TFit}.
}
\end{figure}
%%%%%%%%%%%%%%%%%%%%%%%%%%%%%%%%%%%%%%%%%%%%%%%

  The model parameters have been determined from the fit of the $\pi\pi$ 
scattering phase $\delta_0^0$ and the inelasticity parameter $\eta_0^0$ in the
mass range $M_{\pi\pi}\leq 1.5\;$GeV, see Table~\ref{TFit} and
Figs.\ref{Fit}(a,b). Since the fit was found to be only weakly
sensitive to the form--factor parameter $\lambda_2$, its value was fixed, 
and only the coupling constants $g_{ij}$, the bare mass of $q\bar{q}$, 
and $\lambda_1$ were treated as free parameters. The parameter $b$ in 
the diagonal $\pi\pi$ potential Eq.(\ref{v11}) was used for fine
tuning of the $\pi\pi$ scattering length
which was fixed at $a_0^0 = 0.22 m_{\pi}^{-1}$.
  The best fit of only the $\pi\pi$ scattering data (fit 1) does not lead
automatically to a very good description of the $\pi\pi$ invariant
mass distribution in the decay $\phi\to\gamma\pi\pi$.
However, the shape of the $\phi\to\gamma\pi\pi$ improves
if the $d\Gamma/dM_{\pi\pi}$ data are added to the fit
as shown in Fig.\ref{FitGamma}
(the details are discussed in Sec.\ref{DACCM}).
As a result, our model provides a good description of the whole data set.

  The energy dependence of the $T$-matrix element
$t_{K^+K^\to\pi\pi}$ calculated in our coupled channel model is
shown in Fig.\ref{FigT12}. In addition to a narrow peak due to the
$f_0(980)$ state, this matrix element has a significant
contribution from the lower mass region corresponding to the
$\sigma$ meson.

%%%%%%%%%%%%%%%%%%%%%%%%%%%%%%%%%%%%%%%%%%%%%%%%%%%%%%%%%%%%%%%%%%%%%%%%%%%
\section{The Decay $\phi\to\gamma\pi\pi$ in the Coupled Channel Model}
\label{DACCM}
%%%%%%%%%%%%%%%%%%%%%%%%%%%%%%%%%%%%%%%%%%%%%%%%%%%%%%%%%%%%%%%%%%%%%%%%%%%

%%%%%%%%%%%%%%%%%%%%%%%%%%%%%%%%%%%%%%%%%%%%%%%
\begin{figure}[ht]
\begin{center}
\mbox{(a) \hspace{45mm} (b) \hspace{45mm} (c) \hspace*{20mm}}\\[-5mm]
\mbox{
\mbox{\epsfxsize=45mm \epsffile{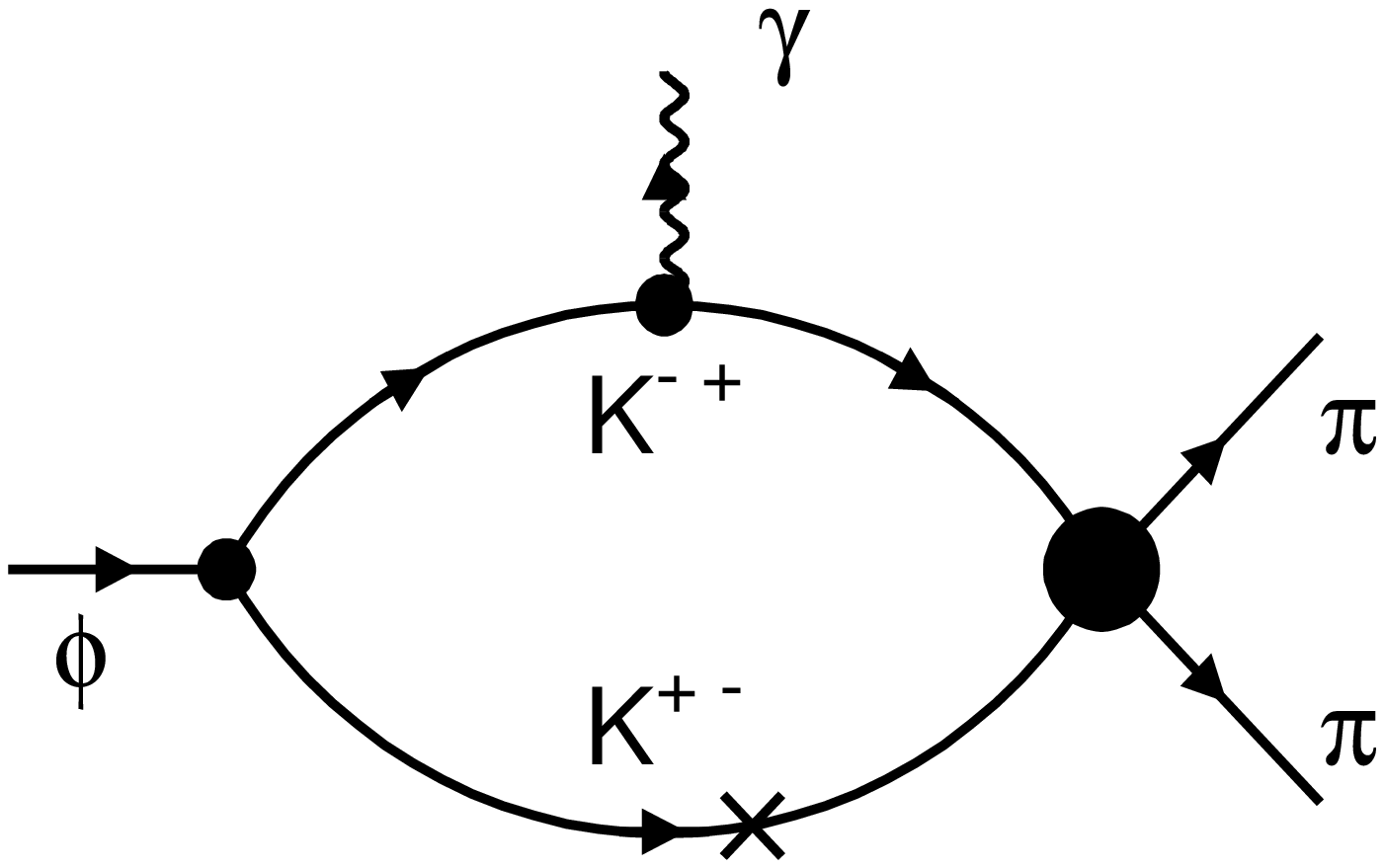}} \hspace{5mm}
\mbox{\epsfxsize=45mm \epsffile{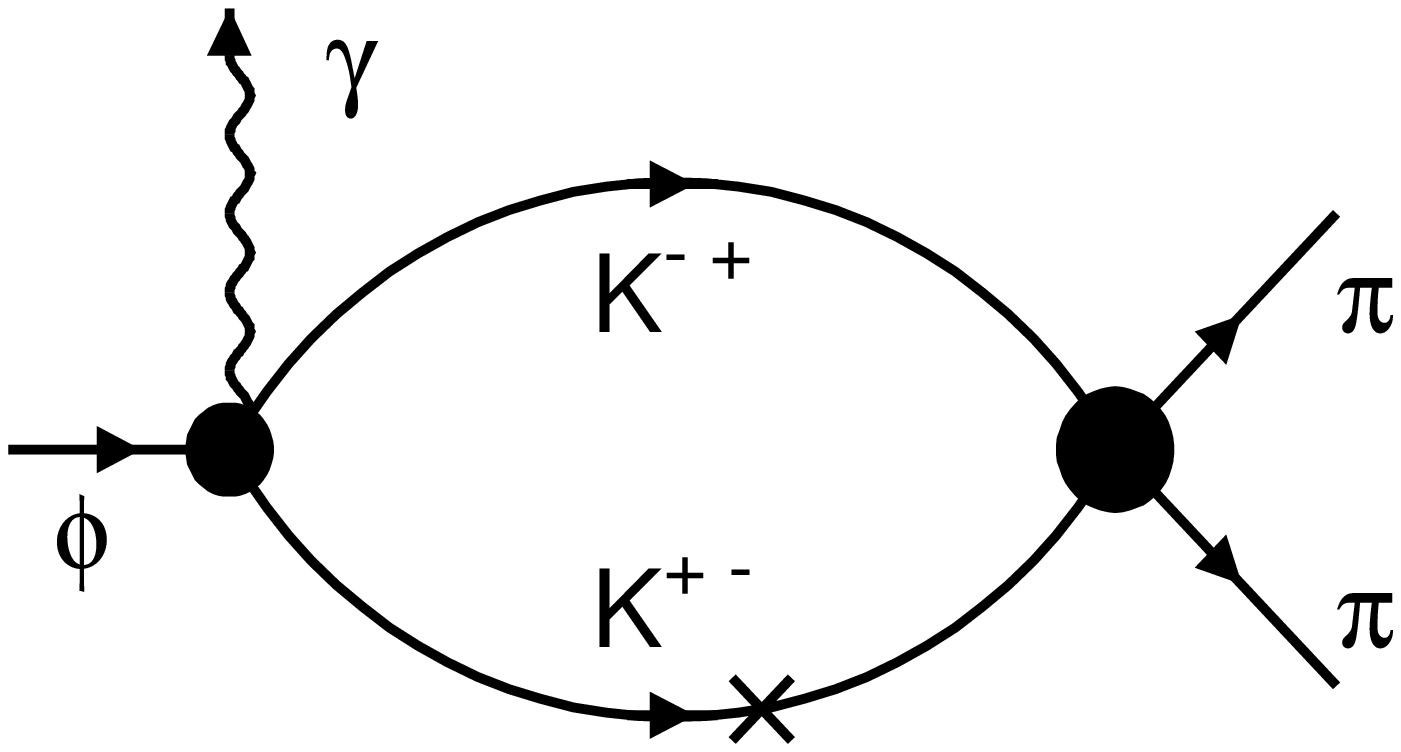}} \hspace{5mm}
\mbox{\epsfxsize=45mm \epsffile{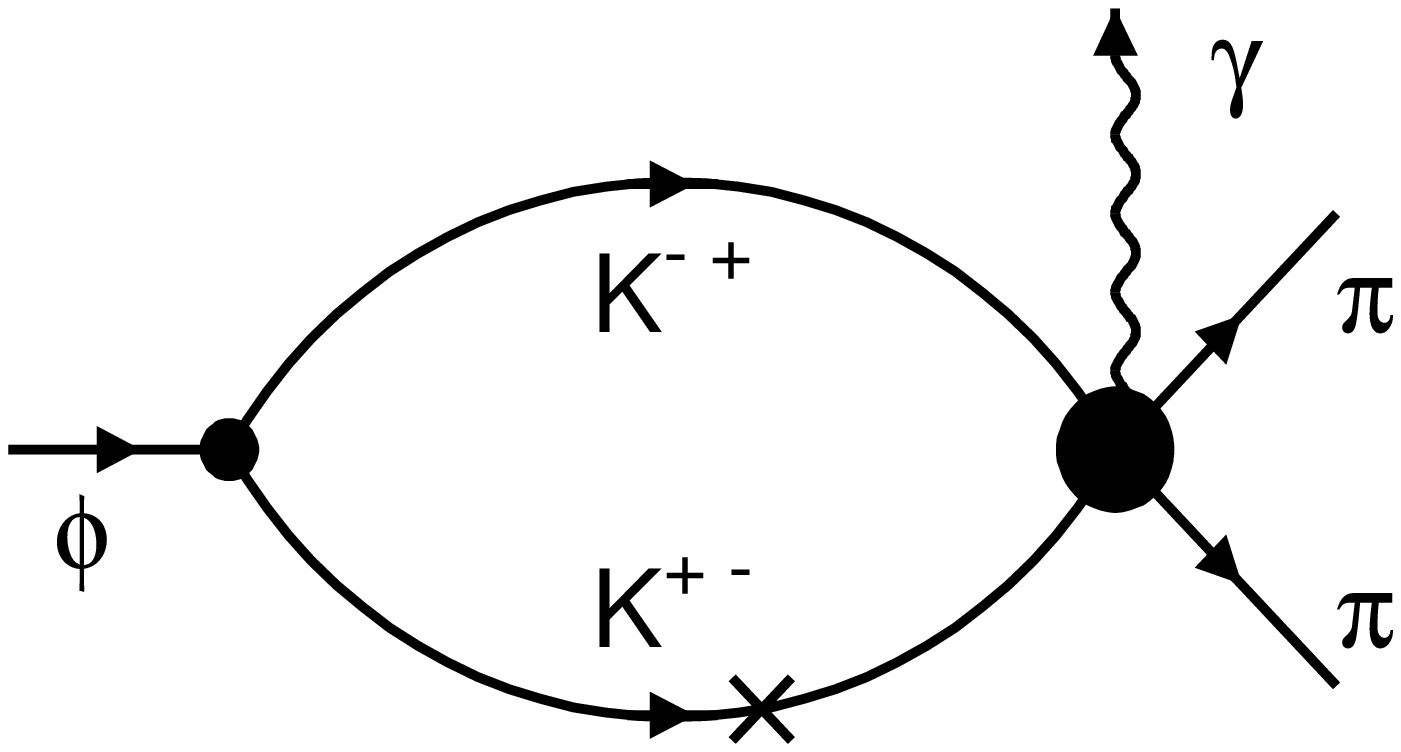}}
}
\caption{\label{FigMphigpipi}%
The minimal set of diagrams giving a gauge-invariant amplitude
$M(\phi\to\gamma \pi\pi)$ in the coupled channel model:
(a) the loop radiation; (b) the contact (seagull) term,
(c) the form factor term.
In the nonrelativistic approximation, the particles marked with $\times$
are on mass shell.
}
\end{center}
\end{figure}
%%%%%%%%%%%%%%%%%%%%%%%%%%%%%%%%%%%%%%%%%%%%%%%

%%%%%%%%%%%%%%%%%%%%%%%%%%%%%%%%%%%%%%%%%%%%%%%
\begin{figure}[htb]
\begin{center}
\mbox{(a) \hspace{80mm} (b)} \\[-5mm]
\mbox{
\mbox{\epsfxsize=70mm \epsffile{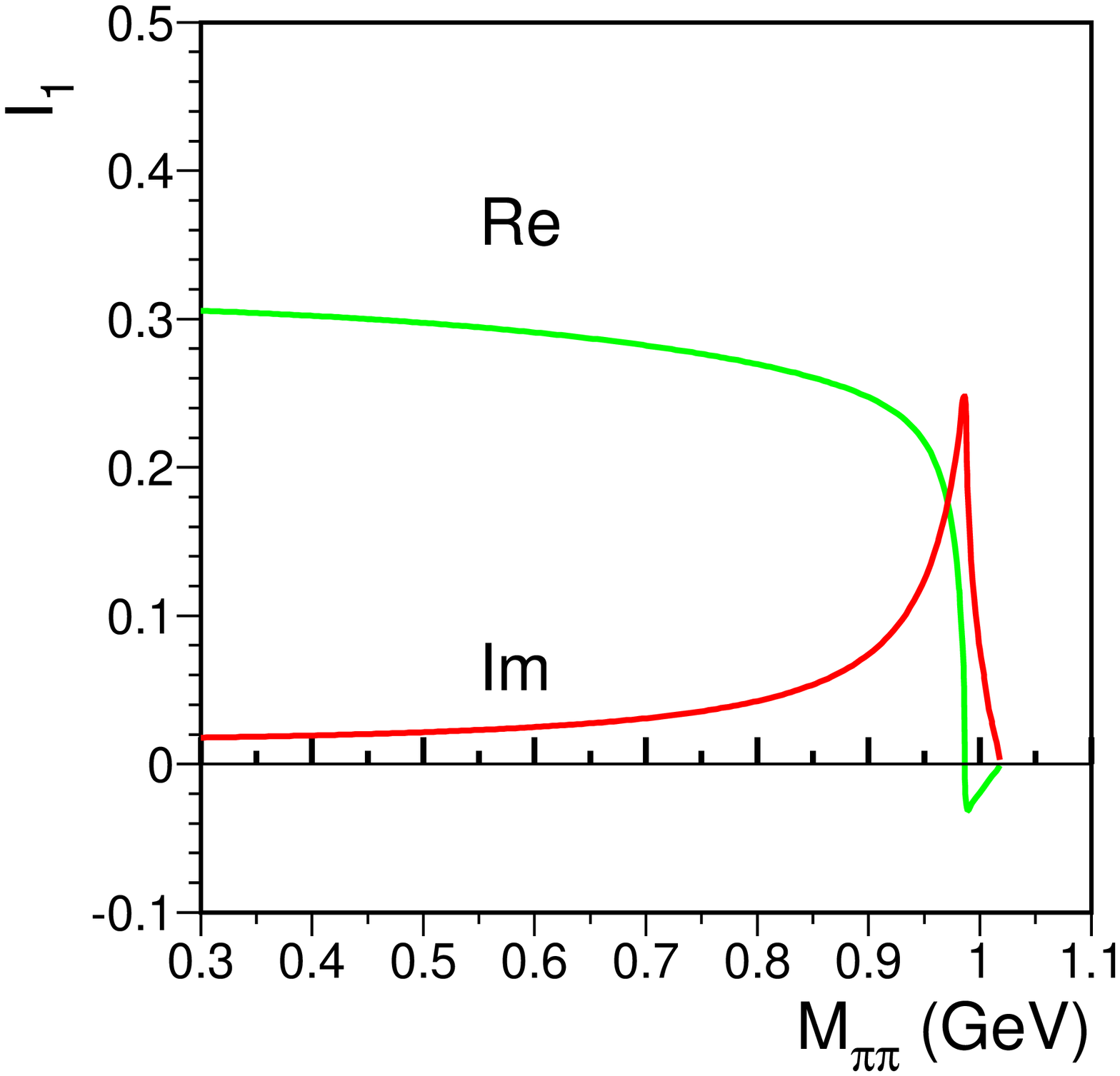}}
\mbox{\epsfxsize=70mm \epsffile{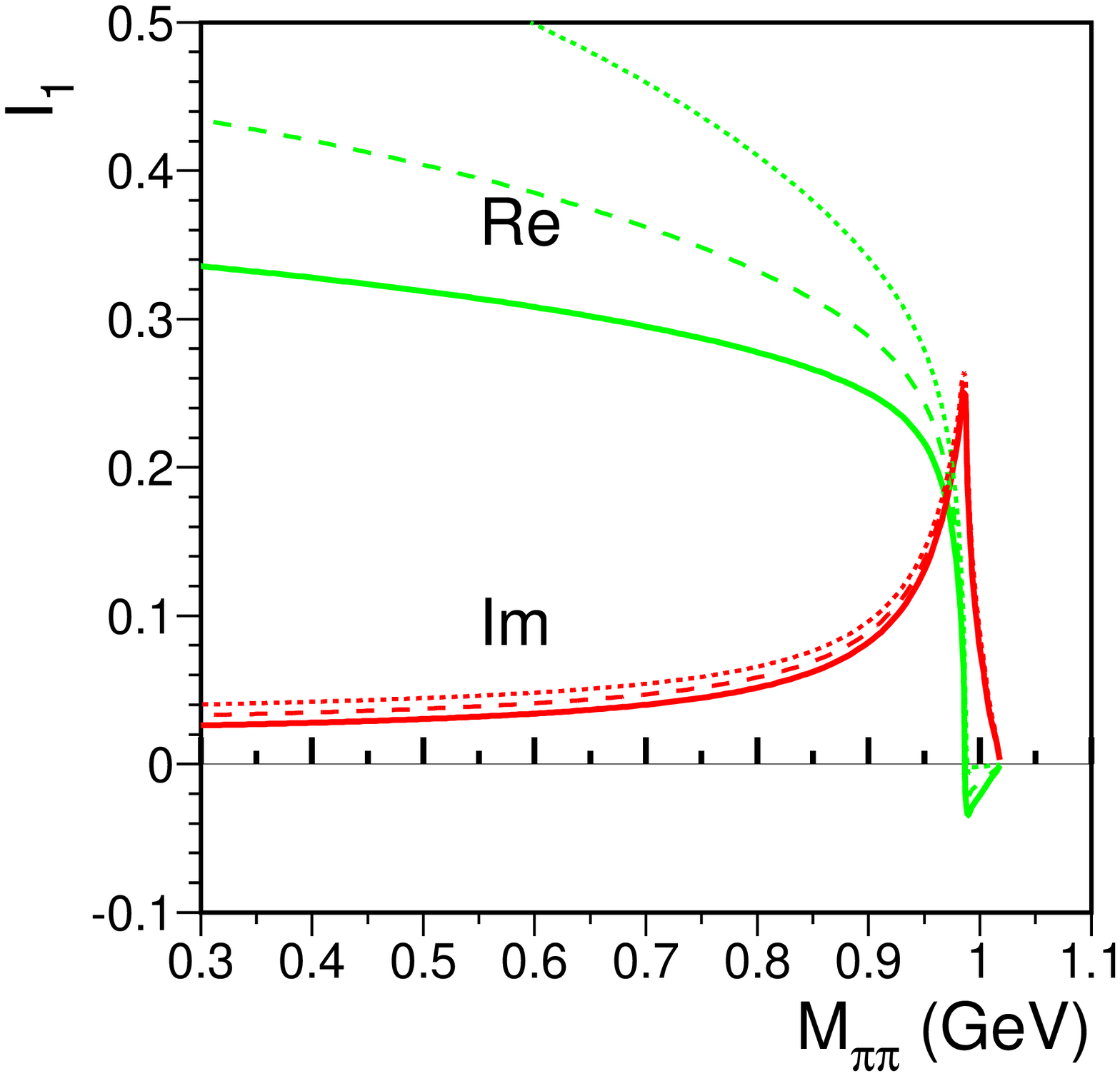}}
}
\end{center}
\caption{\label{FigIJ}%
(a) The function $I_1=(a-b)I(a,b)$, $a=m_{\phi}^2/m_K^2$,
$b=M_{\pi\pi}^2/m_K^2$ vs. the invariant mass of the $\pi\pi$
system $M_{\pi\pi}$. \quad (b) Its nonrelativistic analog  
\mbox{$J_{\lambda}(M_{\pi\pi})\cdot(m_{\phi}^2-M_{\pi\pi}^2)/m_K^2$} for
$\lambda=0.5\;$GeV (solid lines),
$\lambda=0.6\;$GeV (dashed lines),
and $\lambda=0.8\;$GeV (dotted lines).
%The dash-dotted line {\it (to be done)} shows the nonrelativistic result
%obtained from the Siegert-like transformation of the triangle diagram in
%Fig.~\protect\ref{FigMphigpipi}(a).
}
\end{figure}
%%%%%%%%%%%%%%%%%%%%%%%%%%%%%%%%%%%%%%%%%%%%%%%

The formulas given in Sec.~\ref{DA} are valid for point-like
particles. When a form factor in the $K\bar{K}-\pi\pi$ vertex is
included, the $\phi\to\gamma\pi\pi$ amplitude contains an
additional term arising from the minimal substitution $k_{\mu} \to
(k_{\mu} - eA_{\mu})$ in the momentum dependence of the form
factors (see \cite{CIK93} for details).  With an appropriate choice
of the form factors, the sum of three diagrams shown in
Fig.~\ref{FigMphigpipi} becomes explicitly finite. In order to
study the influence of the form factors on the results for the
$\phi\to\gamma\pi\pi$ we use a nonrelativistic approximation for
the $K^+$ and $K^-$, which is justified by the fact that the most
interesting region corresponding to the $f_0(980)$ resonance is
very close to the $K\bar{K}$ threshold.  The electric dipole matrix
element is factorized into two parts describing the $K^+K^-$ loop
radiation with gauge invariant complement and the final
state rescattering $K^+K^-\to\pi\pi$, correspondingly. The total
scalar invariant amplitude has the form
\begin{eqnarray}
 H_{\pi\pi}^{(\lambda)}(p^2,q^2) & = &
 \frac{e g_{\phi}}{2\pi^2 m_K^2}
 J_{\lambda}(M_{\pi\pi}) t_{K^+K^\to\pi\pi}(M_{\pi\pi})
\label{HphigpipiCCM}
\end{eqnarray}
which is similar to the relativistic point-like case defined
by Eq.(\ref{Hphigpipi}) where the function $I(a,b)$
is replaced by $J_{\lambda}(M_{\pi\pi})$.
For the definition of $J_{\lambda}(M_{\pi\pi})$ and further technical
details we refer to \ref{AppNR}.
The parameter $\lambda$ refers to the form--factor
dependence on the relative $K^+K^-$ momentum in the $K^+K^-\to\pi\pi$
vertex given by Eq.(\ref{FF}).  Figure~\ref{FigIJ} shows
the dependence of the electric dipole matrix element on the $\pi\pi$
invariant mass for different values of the form--factor parameter
$\lambda$ in comparison with the relativistic point-like case.

With our choice of the form factor (\ref{FF}) there is no
substantial suppression of the nonrelativistic result\footnote{%
The stronger dependence of the total amplitude on the form factor
found in \cite{CIK93} results from the use of the dipole form
factor which falls off faster than the monopole form factor used in our
case.} in comparison to the point-like case for the relevant range
of $\lambda=0.6-0.8\;$GeV (which corresponds
to the data fit in our coupled channel model). For larger
values of $\lambda$, the full relativistic treatment is needed as
the real part of the nonrelativistic result becomes sensitive to
the short distance contribution (see the discussion in \cite{CIK93}).
The dependence of the imaginary part of $M(\phi\to\gamma\pi\pi)$
on the form factor in the mass region close to the $K\bar{K}$
threshold is rather weak.  As a result, we can neglect the
form--factor dependence in the loop calculations and just use the
relativistic point--like result given by Eq.(\ref{Hphigpipi}).
  It is interesting to note that the imaginary part can be obtained
from the divergent triangle diagram Fig.\ref{FigMphigpipi}(a)
alone by using the Siegert theorem which ensures a correct behaviour
of the electric dipole matrix element.
%%% Appendix?

%%%%%%%%%%%%%%%%%%%%%%%%%%%%%%%%%%%%%%%%%%%%%%%%%%%%%%%%%%%%%%%%%%%%%%%%%%%
\section{The Poles of the $S$-Matrix}
\label{POLES}
%%%%%%%%%%%%%%%%%%%%%%%%%%%%%%%%%%%%%%%%%%%%%%%%%%%%%%%%%%%%%%%%%%%%%%%%%%%

\begin{figure}
\begin{center}
\raisebox{90mm}{(a)}
%%% FitH312y4
\mbox{\epsfysize=110mm\epsffile{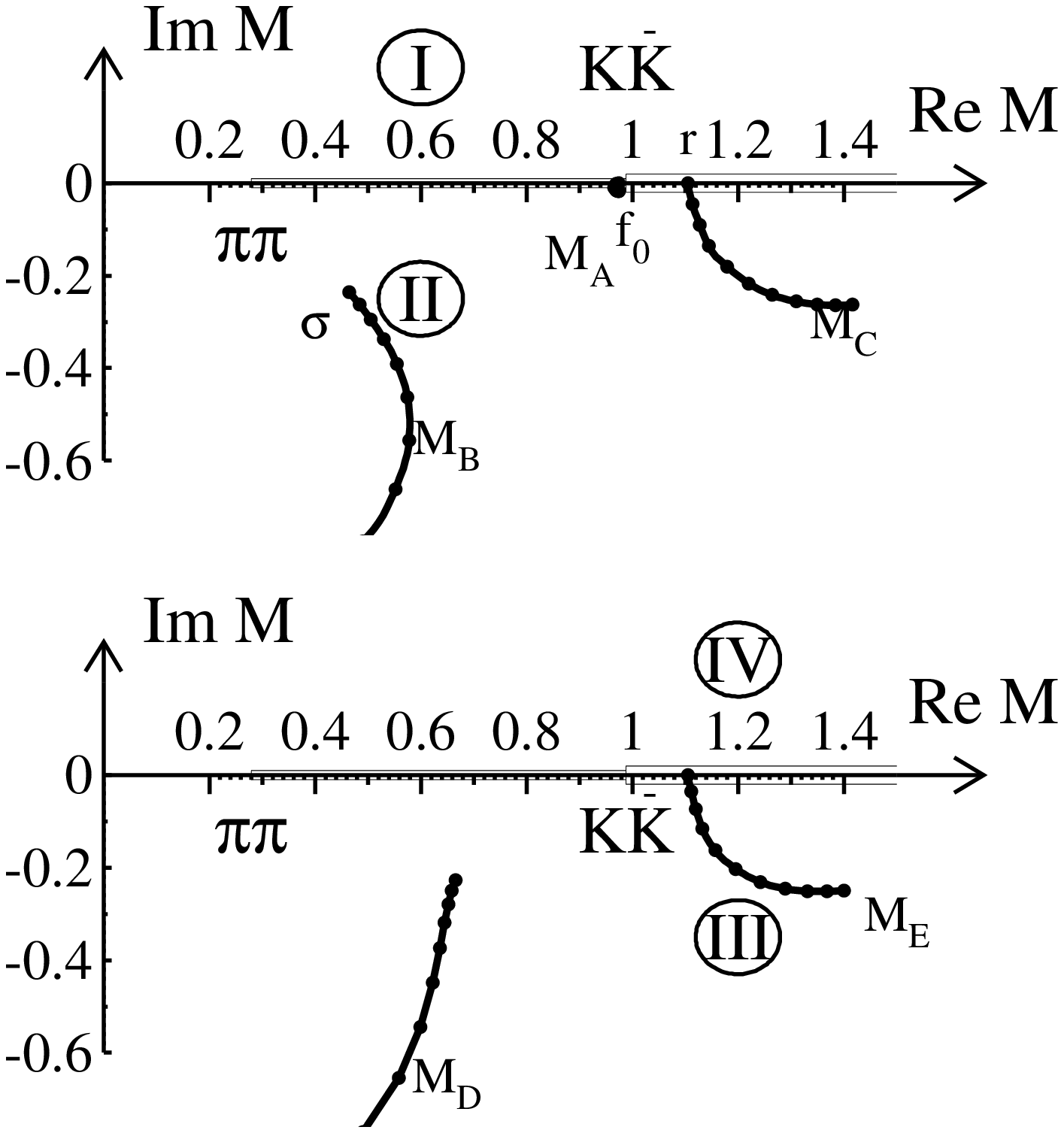}}
\\
\mbox{(b) \hspace{70mm} (c)}\\[-5mm]
\mbox{
%%% FitH312y4
\mbox{\epsfysize=70mm\epsffile{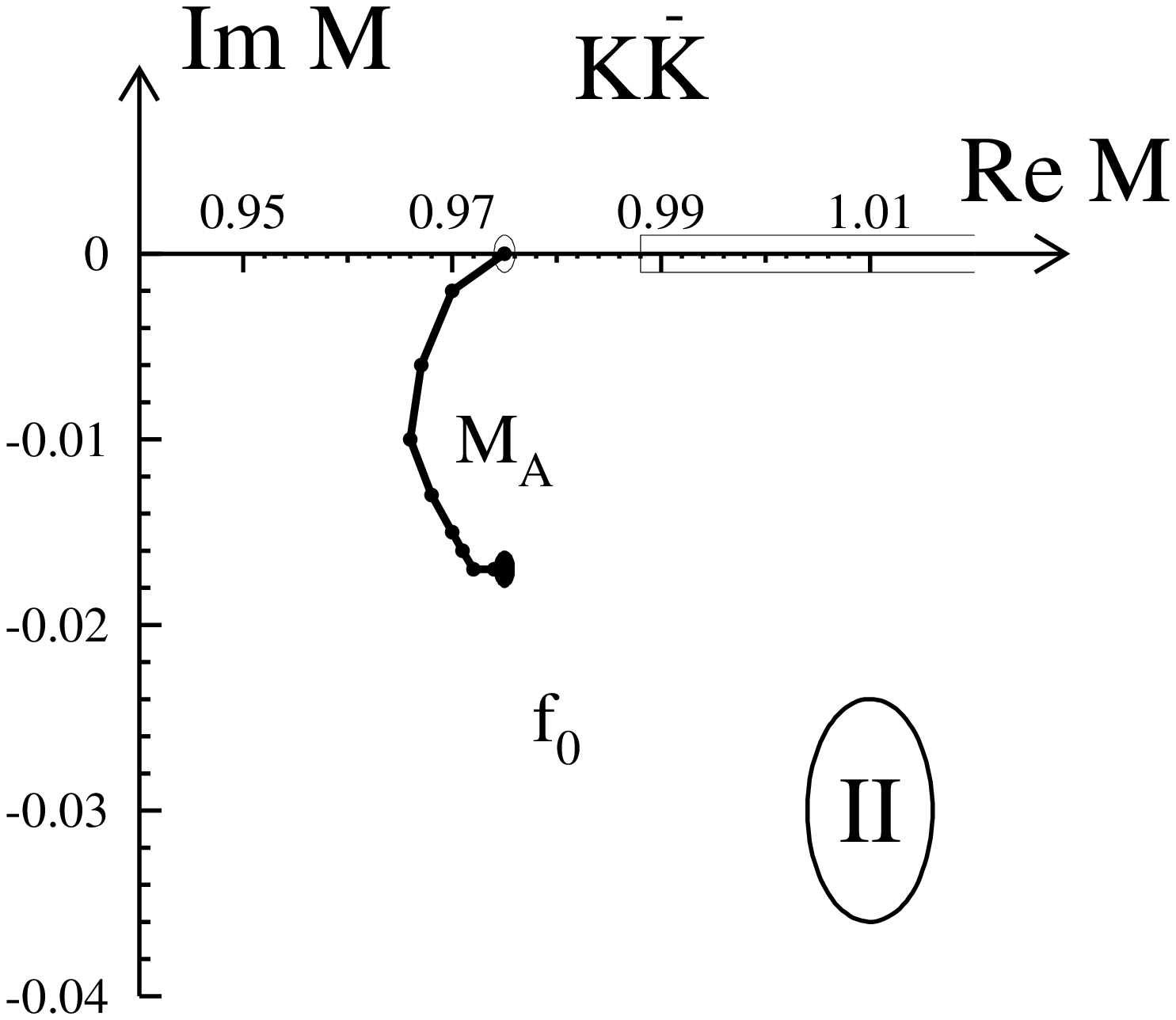}}
\mbox{\epsfysize=70mm\epsffile{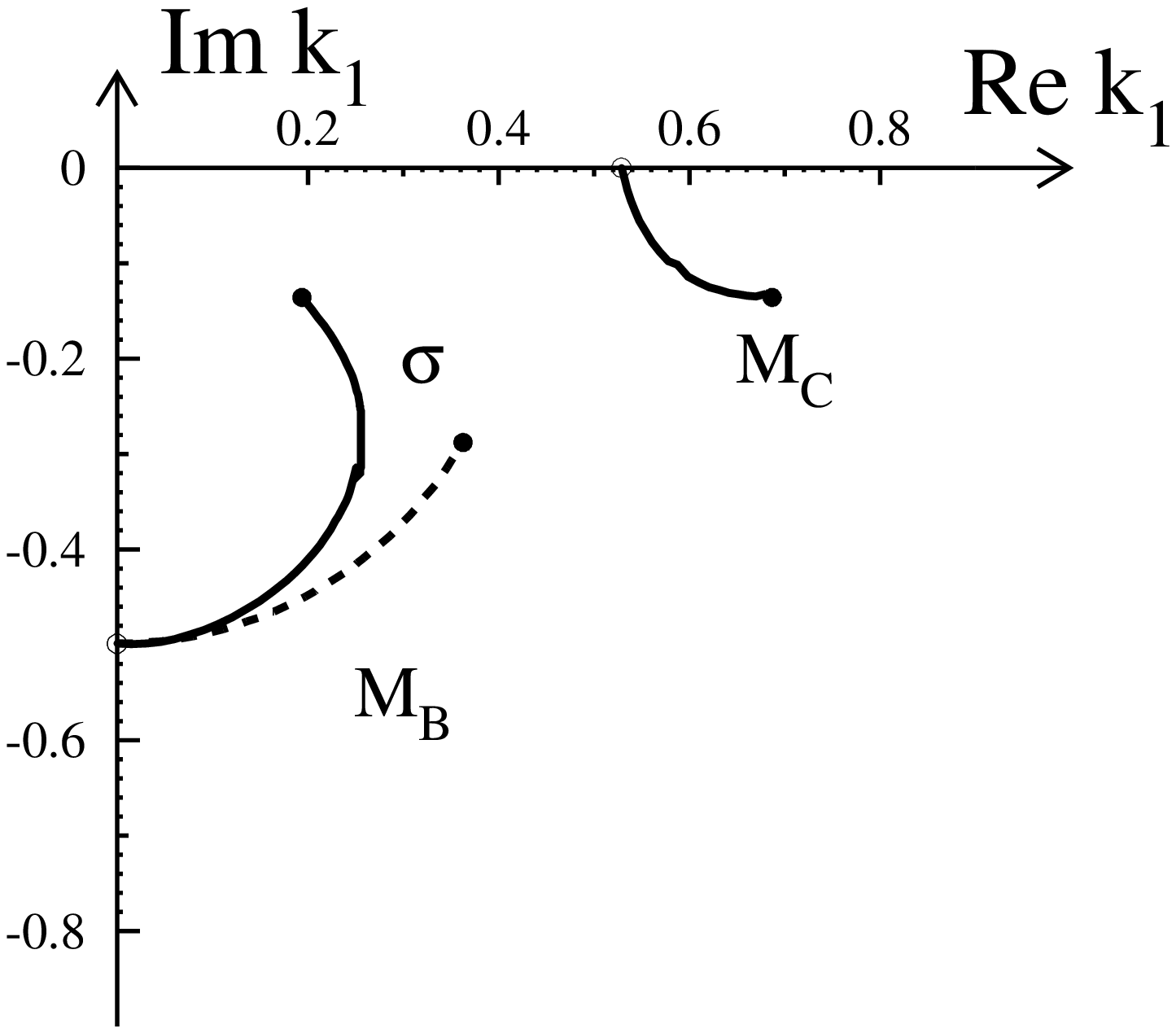}}
}
\end{center}
\vspace*{-5mm}
\caption{\label{FigPoles}
(a) The trajectories of the $S$-matrix poles (fit 4) 
in the complex mass plane (GeV) 
on the sheets II and III  for the $K\bar{K}-\pi\pi$ and $q\bar{q}-\pi\pi$
couplings increasing from $x=0$ to the physical values $x=1$.
%($\circ$) ($\bullet$).
%The labels indicate the original positions of the bound state $(b)$,
%the $q\bar{q}$ resonance $(r)$, and the dynamical pole $(d)$.
%The dots on the trajectories mark the increase of $x$ in steps of $0.1$.
(b) The magnified region of the $f_0(890)$ resonance.
(c) The $\sigma$ meson trajectory in the complex plane of the $\pi\pi$ relative
momentum $k_1$ (GeV) with (solid line) and without (dashed line)
coupling to the $q\bar{q}$ state.
}
\end{figure}

The poles of the $S$-matrix in the complex $s$-plane corresponding to
the fits in Sect.\ref{CCM} are shown in Table~\ref{TPole} and Fig.\ref{FigPoles}.
There are five poles related to the resonances in our model.
The pole $M_A$ located on the sheet II
(${\mathrm Im}\;k_1<0$, ${\mathrm Im}\;k_2>0$)
is very close to the $K\bar{K}$ threshold.
The pair of poles, $M_B$ on sheet II and $M_D$ on the sheet III
(${\mathrm Im}\;k_1<0$, ${\mathrm Im}\;k_2<0$), corresponds to a broad
structure associated with the $\sigma$ meson.
The other pair of poles, $M_C$ on sheet II and $M_E$ on sheet III,
corresponds to a broad resonance above the $K\bar{K}$ threshold,
which can be associated with the $f_0(1370)$ state (for a more realistic
description of the $S$-matrix above 1.3~GeV additional poles are needed
which are not included in the present model).
  The position of the resonance poles depends on the coupling
constants, and this distinguishes them from the fixed poles originating
from the singularities of the form factors (\ref{FF}). The latter are located
at $k_1=\pm i\lambda_1$ and $k_2=\pm i\lambda_2$, their distance to the physical
region being determined by the range of the interaction. In our model
these fixed poles approximate the potential singularities which correspond
to the left hand cut in a more general case.

%%%%%%%%%%%%%%%%%%%%%%%%%%%%%
\begin{table}[hbt]
\caption{\label{TPole}%
The resonance poles of the $S$-matrix in the complex mass plane $M=\sqrt{s}$ (GeV).
The definition of the fits is given in Table~\protect\ref{TFit}.
}
\begin{center}
\begin{tabular}{c|c|c|c|c|c}
\hline
   Pole    & Sheet
   &  fit 1            &  fit 2           &  fit 3            &  fit 4
%% FitH312z4           FitH312y4          FitH312c            FitH312A
\\
\hline
   $M_A$   & II    &
     $0.991 - i0.024$ & $0.974 - i0.020$ & $0.978 - i0.015$ & $0.975 - i0.017$
\\
   $M_B$   & II    &
     $0.455 - i0.237$ & $0.456 - i0.232$ & $0.469 - i0.234$ & $0.465 - i0.236$
\\
   $M_C$   & II    &
     $1.214 - i0.262$ & $1.401 - i0.266$ & $1.424 - i0.261$ & $1.417 - i0.263$
\\
   $M_D$   & III   &
     $0.789 - i0.487$ & $0.677 - i0.234$ & $0.670 - i0.266$ & $0.666 - i0.227$
\\
   $M_E$   & III   &
     $1.403 - i0.320$ & $1.384 - i0.261$ & $1.408 - i0.250$ & $1.400 - i0.249$
\\
\hline
\end{tabular}
\end{center}
\end{table}
%%%%%%%%%%%%%%%%%%%%%%%%%%%%%

%%%%%%%%%%%%%%%%%%%%%%%%%%%
%%%%%%%%%%%%%%%%%%%%%%%%%%%

The origin and the nature of the resonance poles found in our model
can be elucidated by studying how these poles move in the complex $s$-plane
when the model parameters are varied between the physical case
determined by the fit and the limit of vanishing couplings in the 
$\pi\pi$ channel and between the $\pi\pi$ and the other channels
($K\bar{K}$ and $q\bar{q}$):
\begin{eqnarray}
  v_{11}(s) & \to & x \cdot v_{11}(s) \quad , \quad  0 \leq x \leq 1
\\
  v_{12}(s) & \to & x \cdot v_{12}(s) 
\\
  g_{13}    & \to & x^{1/2} \cdot g_{13}
\\
  g_{23}    & \to & x^{1/2} \cdot g_{13}
\end{eqnarray}
The diagonal interaction in the $K\bar{K}$ channel with the {\it
physical} strength of the coupling $g_{22}$  produces a bound state close to
the $K\bar{K}$ threshold with mass $m_{K\bar{K}}=0.97\;$GeV.

%%%%%%%%%%%%%%%%%%%%%%%%%%%
\begin{figure}[htb]
\begin{center}
\mbox{(a) \hspace{80mm} (b) \hspace{30mm}}\\[-5mm]
\mbox{
%%% Update 10/05/2000 FitH312y4:
\hspace*{-5mm}
\mbox{\epsfxsize=80mm\epsffile{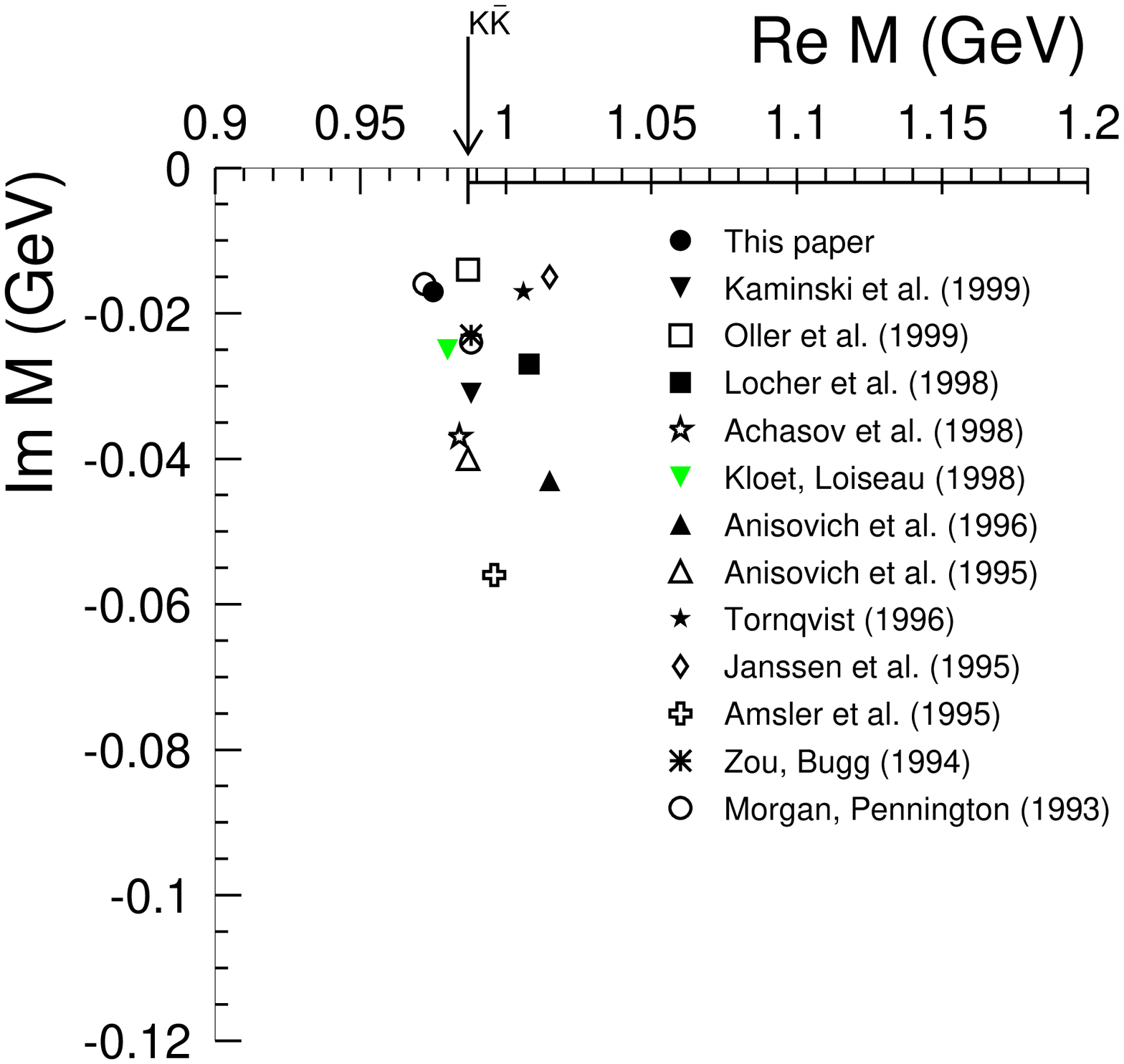}}
\mbox{\epsfxsize=80mm\epsffile{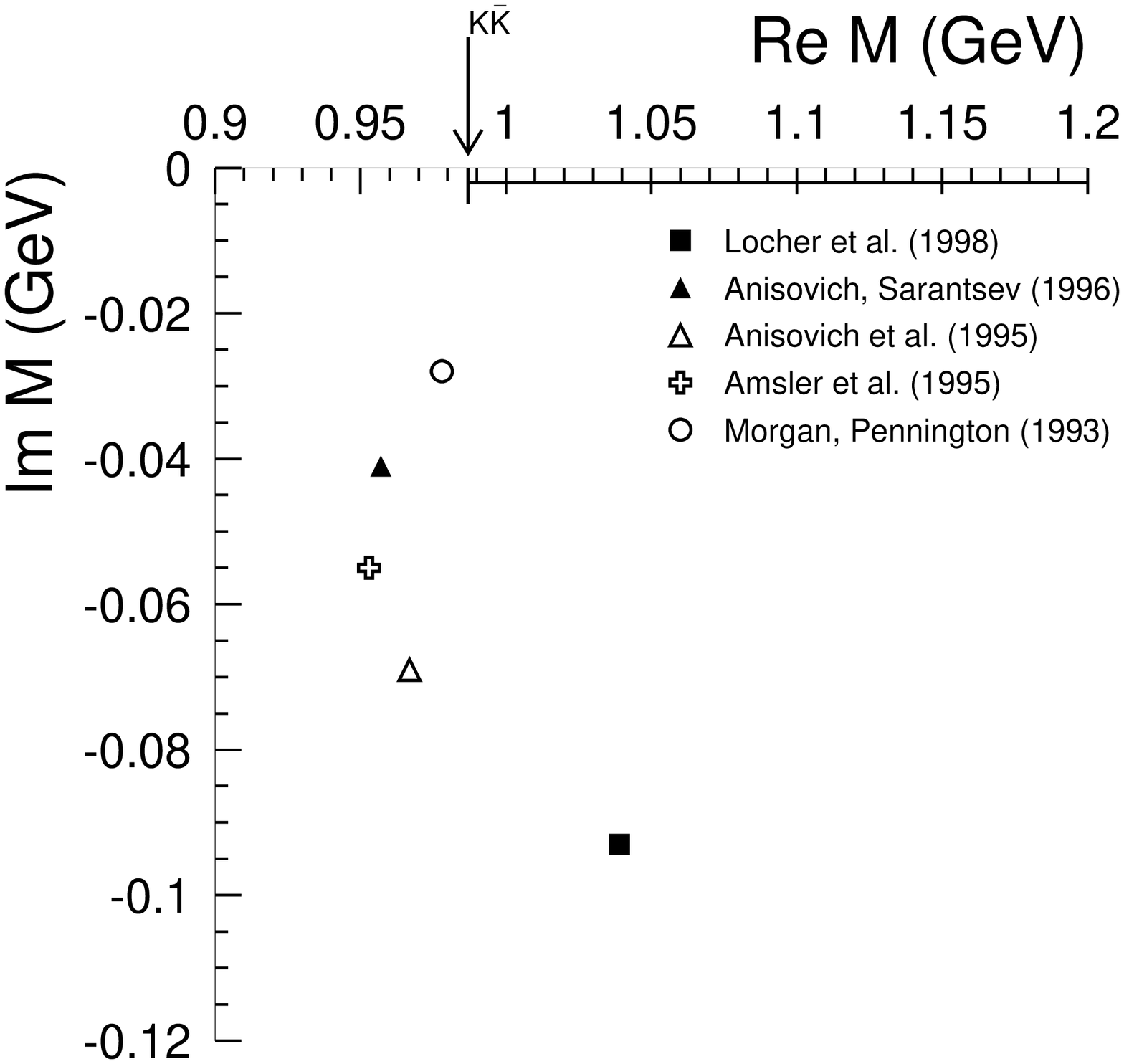}}}
% > perl plotTab2.pl < f0s2Tab*.dat > f0s2Tab*.kumac
% PAW> exec f0s2Tab000510.kumac
% PAW> pic/pri Plotf0s2Tab.eps
\end{center}
\caption{\label{FigSumf0}%
The poles of the $S$-matrix corresponding to the $f_0(980)$
in different models
\cite{SND98,KLL99,LMZ98,JPHS95,To96,OO99,MP93,ZB94,CBCg,An96,AS96,KL98}:
(a) sheet II pole ($\mbox{Im}\;k_1>0$, $\mbox{Im}\;k_2<0$) and
(b) sheet III pole ($\mbox{Im}\;k_1<0$, $\mbox{Im}\;k_2<0$).
}
\end{figure}
%%%%%%%%%%%%%%%%%%%%%%%%%%%

  Our coupled channel model has only {\it one} pole $M_A=0.975-i0.017\;$GeV
near the $K\bar{K}$ threshold, which is sufficient for a good description of
the $\pi\pi$ scattering data.  This pole is directly related to a molecular
$K\bar{K}$ state in the absence of coupling to the $\pi\pi$ channel.
  The number of the $S$-matrix poles near the $K\bar{K}$ threshold has been
discussed in the literature (see \cite{LMZ98,MP93,AMP87} and references therein)
for a long time.
While the relation of the $f_0(980)$ meson to at least one $S$-matrix pole
close to $K\bar{K}$ threshold is well established,
the exact location and even the number of the relevant $S$-matrix poles
is model dependent as demonstrated in Fig.\ref{FigSumf0}.
  The models based on dynamical input (coupled channel,
potential, unitarized chiral perturbation theory)
\cite{CDL89,KLM94,KLL99,JPHS95,To95,OO97,Ol98}
usually produce only one pole which can be traced to a weakly
bound $K\bar{K}$ state in an appropriate limit of the channels coupling.
  One exception to this observation is a coupled channel model \cite{LMZ98} where
the second pole near the $K\bar{K}$ threshold results from the interplay
of a nearby $q\bar{q}$ pole with dynamical singularities.
  We were not able to find a good fit with the same feature in our model;
the main reason appears to be due to the using of the energy dependent potentials
in Eq.(\ref{Vsep}) contrary to the case in \cite{LMZ98}.  However, it is not
excluded that two--pole solutions can be found with some other parametrization
of interactions in CCM.  It remains to be investigated whether two--pole
solutions are compatible with the data on the $\phi\to\gamma\pi\pi$ decay.

  The $K$-matrix parameterizations \cite{MP93,AMP87,ZB94,ABSZ94,CBCg,An95,An96,AS96}
routinely produce the second pole on sheet III. There is, however, a
much larger spread in this pole location than on sheet II.
As shown above, the reaction $\phi\to\gamma\pi\pi$ allows one to probe
the scattering $K\bar{K}\to\pi\pi$ both above and below the $K\bar{K}$
threshold, and the $\pi\pi$ mass distribution in $\phi\to\gamma\pi\pi$
is very sensitive to the $S$-matrix poles related to the $f_0(980)$. Therefore
more detailed experimental data on $\phi\to\gamma\pi\pi$ would be very useful
for reducing the present uncertainty about the analytical structure
of the $S$-matrix in the $f_0(980)$ region.

%%%%%%%%%%%%%%%%%%%%%%%%%%%%%%%%%%%%%%%%%%%%%%%%%%%%%%%%%%%%%%%%%%%%%%%%%%%
\section{The Mixing between the $q\bar{q}$ and Mesonic Channels}
\label{Mixingqqbar}
%%%%%%%%%%%%%%%%%%%%%%%%%%%%%%%%%%%%%%%%%%%%%%%%%%%%%%%%%%%%%%%%%%%%%%%%%%%

%%%%%%%%%%%%%%%%%%%%%%%%%%%
\begin{figure}[htb]
\begin{center}
\mbox{(a) \hspace{70mm} (b)}
\mbox{
\mbox{\epsfysize=70mm \epsffile{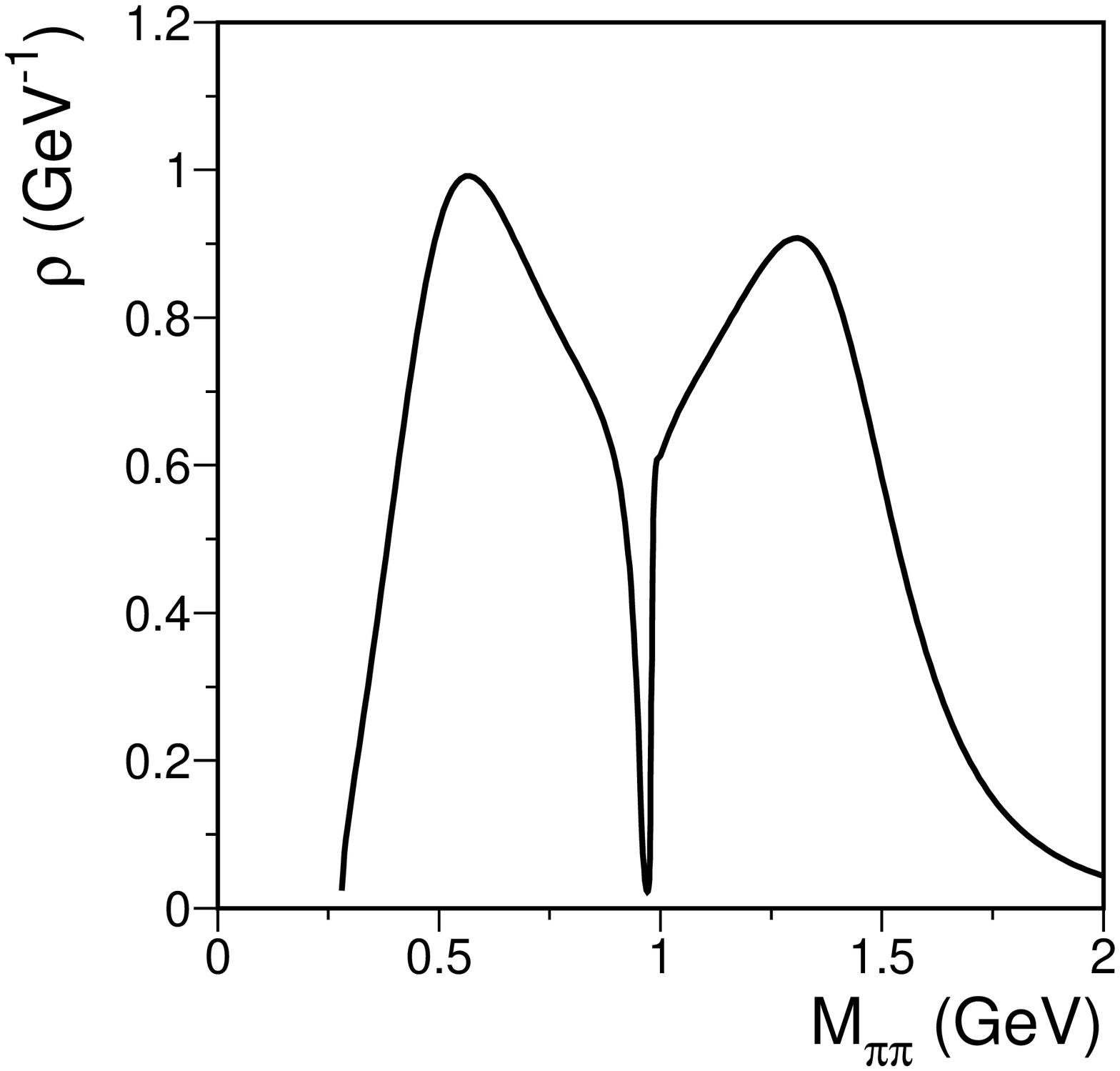}}
\mbox{\epsfysize=70mm \epsffile{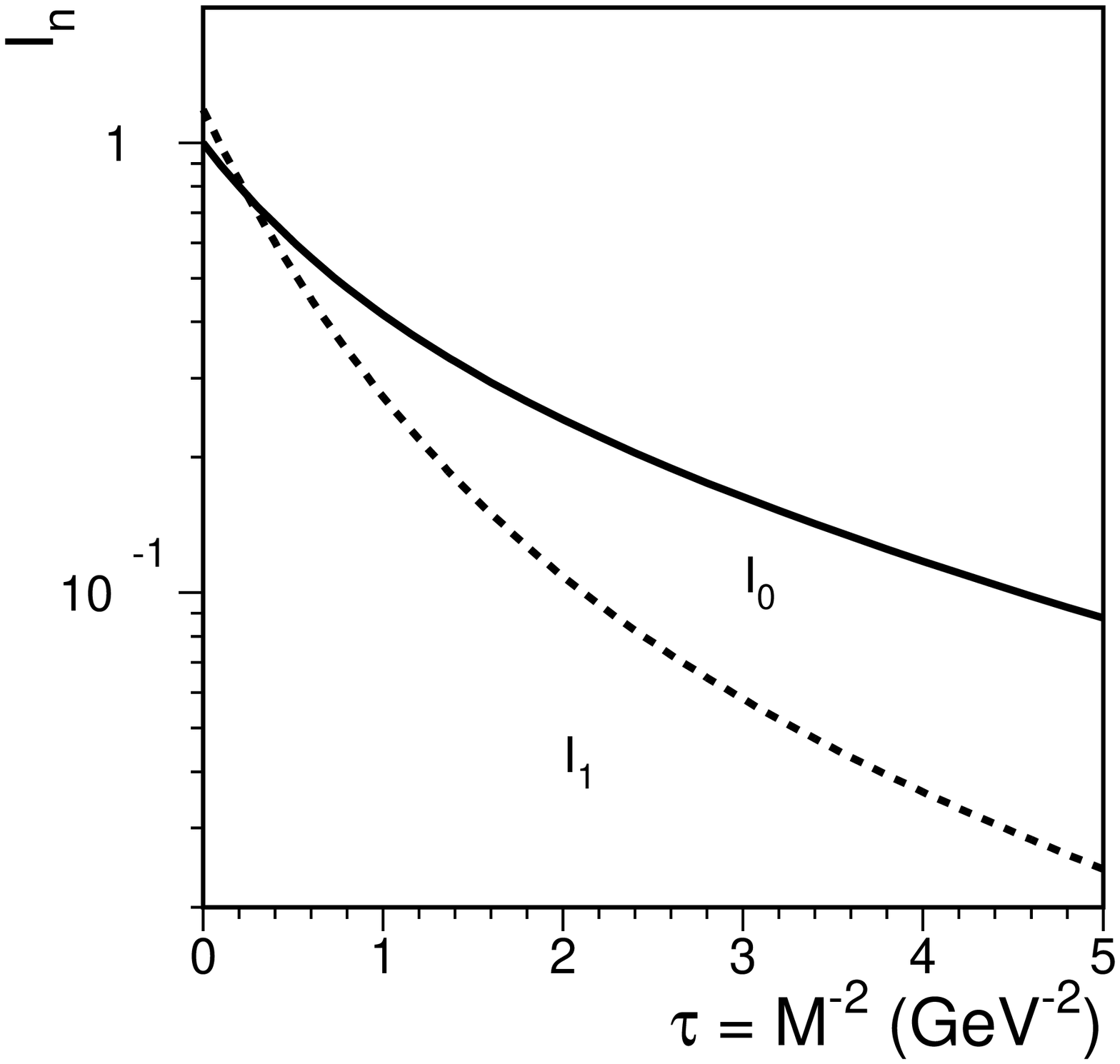}}
% CCM00Hfit_poles.MCD (09/05/00) => I0I1MH312y4.dat
% PAW> exec FigLaplaceSD.kumac '../MCD/I0I1MH312y4.dat' [Color]
% PAW> pic/pri PlotLaplaceSD.eps
}
\end{center}
\caption{\label{SD}%
(a) The spectral density of the $q\bar{q}$ state and
(b) the corresponding Laplace transforms.
}
\end{figure}
%%%%%%%%%%%%%%%%%%%%%%%%%%%

  The mixing of the quark-antiquark states with the open meson channels
can be studied in the CCM by using the probability sum rule for
a resonance embedded into a continuum \cite{BHM} which
is described in \ref{AppCCM}.
The spectral function $\rho(s)$ which determines the probability
density for the $q\bar{q}$ component in the scattering states,
as defined by Eq.(\ref{SumR}), is shown in Fig.\ref{SD}(a).
In the case of weak coupling, the probability density would be well localized near
the position of the bare $q\bar{q}$ state at $M_r=1.1\;$GeV.
For the physical case, we find a
broad peak centered above the $K\bar{K}$ threshold in the region corresponding
to the $f_0(1370)$ resonance (the poles $M_E$ and $M_C$).
There is also a sizable contribution to
the $q\bar{q}$ spectral density from the low--mass region of the $\sigma$ meson.
The $\rho(s)$ distribution in the $f_0(980)$ resonance has a characteristic
dip-bump structure resulting from an interplay of the resonance pole
$M_A$ with a nearby zero of the mass operator $\Pi(s)$.
  The position and the width of the $\rho(s)$ distribution indicates that an
essential contribution to the saturation of the sum rule (\ref{SumR}) comes
from the pole $M_B$ related to the $\sigma$ meson,
while the narrow structure associated with the pole $M_A$ alone plays a minor role.
  The fact that $q\bar{q}$ coupling with the $\pi\pi$ channel significantly
enhances the spectral density $\rho(s)$ in the region of the $\sigma$ meson
is related to the interplay of the $S$-matrix poles demonstrated in
Fig.\ref{FigPoles}c: the pole $M_B$ corresponding to the $\sigma$ meson
is pushed towards the $\pi\pi$ threshold by the pole $M_C$ originating
from the $q\bar{q}$ state.

  Using the spectral density $\rho(s)$ we can calculate the contribution
of the $q\bar{q}$ scalar mesons to the QCD sum rule related to the scalar
quark condensate \cite{SVZ}.  Figure~\ref{SD}b shows the Laplace transform
of the spectral density which is used in the sum rule analysis:
\begin{eqnarray}
  I_n(M^2) & = &
  \int_0^\infty  s^n e^{-s/M^2} \rho(s) ds
\label{LSR}
\end{eqnarray}
The advantage of CCM with respect to the previous studies (see \cite{QCDSR}
and references therein)
where the contribution of the $q\bar{q}$ mesons was usually approximated
by one narrow resonance is a more realistic shape of
the $q\bar{q}$ spectral density in the low-mass region which
is important in the Laplace sum rules.  As a result,
the momenta $I_0(M^2)$ and $I_1(M^2)$ have quite different slopes in
their $M^2$--dependence as shown in Fig.\ref{SD}(b).

%%%%%%%%%%%%%%%%%%%%%%%%%%%%%%%%%%%%%%%%%%%%%%%%%%%%%%%%%%%%%%%%%%%%%%%%%%%
\section{Discussion}  \label{Disc}
%%%%%%%%%%%%%%%%%%%%%%%%%%%%%%%%%%%%%%%%%%%%%%%%%%%%%%%%%%%%%%%%%%%%%%%%%%%

  The theoretical calculations of the decays
$\phi\to\gamma\pi\pi$ and $\phi\to\gamma f_0(980)$
and the corresponding experimental data are summarized in
Table~\ref{TabSum}.
  The predicted branching ratios for the $\gamma\pi^0\pi^0$ and
$\gamma\pi^+\pi^+$ channels should be compared with the experimental
data with some caution.  While these branching ratios can be easily defined
theoretically 
($\Gamma_{\phi\to\gamma\pi^+\pi^-}/\Gamma_{\phi\to\gamma\pi^0\pi^0}=2$ 
for the isoscalar $\pi\pi$ states, e.g.), the data analysis relies on
the modeling of competing reaction mechanisms.
In particular, the interpretation of the experimental
results for the $\gamma\pi^+\pi^+$ channel involves the consideration of
the interference between the $e^+e^-\to \phi\to\gamma\pi\pi$ and
$e^+e^-\to \gamma\rho\to\gamma\pi\pi$ mechanisms.
In this respect, we wish to emphasize again the importance of using 
a realistic $K^+K^-\to\pi\pi$ amplitude instead of a simple
superposition of BW resonances.
The channel $\gamma\pi^0\pi^0$ which is free from the $\rho$ meson 
contribution in the $\pi\pi$ channel appears to be better suited for 
the study of the $\pi\pi$ invariant mass distribution, 
although this case has some background due to the 
mechanism $\phi \to \pi^0\rho^0 \to \pi^0\pi^0\gamma$. 

  Our results are in good agreement with recent calculations
within the chiral unitary approach \cite{MHOT99} both for the total width and
for the $\pi\pi$ mass distribution. This is not surprising
because in both cases the $f_0(980)$ resonance is produced
mainly by the attractive interaction in the $K\bar{K}$ channel.
  Our result for the total width $\Gamma_{\phi\to\gamma\pi\pi}$ is close to
the earlier calculations of the two--step mechanism with the intermediate
$K^+K^-$ state \cite{CIK93,LP90} which used the BW approximation (\ref{GphigS})
with the coupling constant $g_{f_0K\bar{K}}^2/4\pi = 0.6\;\mbox{\rm GeV}^2$.
In our model we can define an {\it effective} coupling constant
$g_{f_0K\bar{K}}$ by approximating the $K\bar{K}\to\pi\pi$ amplitude
(\ref{T12}) by a BW resonance which leads to slightly higher value of
$g_{f_0K\bar{K}}^2/4\pi = 1.1\;\mbox{\rm GeV}^2$.

  The effect of the form factor in the $K\bar{K}\pi\pi$ vertex was
studied earlier in \cite{CIK93} where a suppression by a factor of about 5
was given as an estimate.  This significant suppression resulted from the
use of a very soft dipole form factor with the characteristic parameter
$\mu=0.14\;$GeV (Eq.(4.14) in \cite{CIK93}) which was suggested by the
study of the $\phi\to\gamma\gamma$ decay \cite{Ba85}. However, the $2\gamma$
decay is related to the short distance behavior of the $K\bar{K}$ wave function,
while we are interested in the $\phi K\bar{K}$ vertex at moderate
relative momenta (in general, there is no unique relation between these
two properties, see the discussion in \cite{Ba85}).  It is difficult
to justify such a low value of the form--factor parameter in the CCM:
a good fit of the scattering data requires much harder form factors
as discussed in Sec.~\ref{CCM}.  A weak effect of the form factor found
in our case is consistent with the results in \cite{CIK93} if
the form--factor parameter $\mu \geq 0.5\;$GeV is used there.

%%%%%%%%%%%%%%%%%%%%%%%%%%%%%%%%%%%%%%%%%%%%%%%%%%%%%%%%%%%%%%%%
\begin{table}
  \centering
  \begin{tabular}{|l|l|c|}
  \hline
    Reaction & Branching ratio & Ref.
  \\
  \hline
  \multicolumn{3}{|c|}{Theory}
  \\ %%%%%%%%%%%%%%%%%%%%%%%%%%%%%%%%%
  \hline
    $\phi\to\gamma f_0(980)\to \gamma \pi\pi$
    & $2.6 \cdot 10^{-4}$ & \cite{AI89}
  \\  %%%%%%%%%%%%%%%%%%%%%%%%%%%%%%%%%
   \hline
    $\phi\to\gamma f_0(980)$
    & $1.9 \cdot 10^{-4}$ & \cite{LP90}  % $8.5 \cdot 10^{-4}$ % see comment by CIK93
  \\  %%%%%%%%%%%%%%%%%%%%%%%%%%%%%%%%%
  \hline
    $\phi\to\gamma f_0(980)$ (point--like)
    & $1.4 \cdot 10^{-4}$  & \cite{CIK93} % $6 \cdot 10^{-4}$
  \\
    $\phi\to\gamma f_0(980)$ (with form factor)
    & $0.3 \cdot 10^{-4}$ & \cite{CIK93}
  \\  %%%%%%%%%%%%%%%%%%%%%%%%%%%%%%%%%
  \hline
    $\phi\to\gamma (f_0+\sigma)\to \gamma\pi\pi$ ($q^2\bar{q}^2$ model)
    & $\sim 10^{-4}$ & \cite{AG97}
  \\
    $\phi\to\gamma (f_0+\sigma)\to \gamma\pi\pi$ ($K\bar{K}$ model)
    & $\sim 10^{-5}$ & \cite{AG97}
  \\
    $\phi\to\gamma (f_0+\sigma)\to \gamma\pi\pi$ ($s\bar{s}$ model)
    & $\sim 5\cdot 10^{-5}$ & \cite{AG97}
  \\  %%%%%%%%%%%%%%%%%%%%%%%%%%%%%%%%%
  \hline
    $\phi\to\gamma (\pi^0\pi^0)_{J=0}$
    & $0.8 \cdot 10^{-4}$ & \cite{MHOT99}
  \\
    $\phi\to\gamma (\pi^+\pi^-)_{J=0}$
    & $1.6 \cdot 10^{-4}$ & \cite{MHOT99}
  \\  %%%%%%%%%%%%%%%%%%%%%%%%%%%%%%%%%
  \hline
    $\phi\to\gamma (\pi^0\pi^0)_{J=0}$
    & $1.2 \cdot 10^{-4}$           & this paper  % FitH312y4
  \\
    $\phi\to\gamma(\pi^0\pi^0)_{J=0}$, ($M_{\pi^0\pi^0}>0.9\;$GeV)
    & $0.34 \cdot 10^{-4}$           & this paper
  \\
    $\phi\to\gamma (\pi^+\pi^-)_{J=0}$
    & $2.3 \cdot 10^{-4}$           & this paper
  \\  %%%%%%%%%%%%%%%%%%%%%%%%%%%%%%%%%
  \hline
  \multicolumn{3}{|c|}{Experiment}
  \\  %%%%%%%%%%%%%%%%%%%%%%%%%%%%%%%%%
  \hline
    $\phi\to\gamma(\pi^0\pi^0)_{J=0}$
    & $(1.14 \pm 0.10 \pm 0.12)\cdot 10^{-4}$ & \cite{SND98}
  \\
    $\phi\to\gamma(\pi^0\pi^0)_{J=0}$, ($M_{\pi^0\pi^0}>0.9\;$GeV)
    & $(0.5 \pm 0.06 \pm 0.06)\cdot 10^{-4}$ & \cite{SND98}
  \\  %%%%%%%%%%%%%%%%%%%%%%%%%%%%%%%%%
    $\phi\to\gamma f_0(980)$
    & $(3.42 \pm 0.30 \pm 0.36)\cdot 10^{-4}$ & \cite{SND98}
  \\
\hline
%    $\phi\to\gamma(\pi^+\pi^-)_{J=0}$
%    & $(0.41 \pm 0.12 \pm 0.04)\cdot 10^{-4}$ ? & \cite{CMD99} %%% ?? check
%  \\
    $\phi\to\gamma f_0(980)$
    & $(1.96 \pm 0.46 \pm 0.50)\cdot 10^{-4}$ & \cite{CMD99a} 
  \\
    $\phi\to\gamma \pi^0\pi^0$
    & $(1.08 \pm 0.17 \pm 0.09)\cdot 10^{-4}$ & \cite{CMD99b}
  \\
    $\phi\to\gamma \pi^0\pi^0$, ($M_{\pi^0\pi^0}>0.9\;$GeV)
    & $(0.57 \pm 0.06 \pm 0.04)\cdot 10^{-4}$ & \cite{CMD99b}
  \\
  \hline
  \end{tabular}
  \caption{\label{TabSum}%
  The calculated branching ratios for $\phi\to\gamma(\pi\pi)_{J=0}$ and
  $\phi\to\gamma f_0(980)$ in comparison with the experimental data.
  }
\end{table}
%%%%%%%%%%%%%%%%%%%%%%%%%%%%%%%%%%%%%%%%%%%%%%%%%%%%%%%%%%%%%%%%

  In view of this overall agreement between different calculations
of the two--step mechanisms with the $K^+K^-$ intermediate state we find
it difficult to agree with the statements \cite{AI89,AG97}
that the $K\bar{K}$ model of $f_0(980)$ can be excluded on the ground
of its alleged conflict with the $\phi\to\gamma f_0(980)$ data.
  Since the hadronic part of the amplitude
$\phi\to\gamma\pi\pi$ is factored out as the matrix element
$t_{K\bar{K}-\pi\pi}$ or, in the simplified case, as the coupling
constant $g_{f_0K\bar{K}}$, the self-consistency
of the $K\bar{K}$ model of $f_0(980)$ can be examined
by analyzing the value of $g_{f_0K\bar{K}}$.
For this purpose we use a well known result from nonrelativistic
scattering theory: the residue of the scattering matrix
at a pole corresponding to a bound state is uniquely related to
the asymptotic normalization constant of the bound state wave function
which for a weakly bound state depends, in leading order, only on the
binding energy.  For a weakly bound $K\bar{K}$ state,
the relation between the $g_{f_0K\bar{K}}$ and
the binding energy $E_b = 2m_K - m_{f_0}$ has the form
\begin{eqnarray}
  \frac{g_{f_0K\bar{K}}^2}{4\pi} & = &
  32 m_K \kappa ( 1 + O(\kappa/\lambda) )
  \label{gfKKB}
\\
  \kappa & = & \sqrt{m_K E_b}
\end{eqnarray}
where $\lambda$ is the range of the $K\bar{K}$ interaction.
% $\kappa/\lambda \ll 1$ for a weakly bound state.
Taking $E_b=4\;$MeV and neglecting small corrections of the
order $\kappa/\lambda$ one gets
$\frac{g_{f_0K\bar{K}}^2}{4\pi}=0.7\;\mbox{\rm GeV}^2$
which is very close to the values discussed above.  Therefore a large
$K\bar{K}$ component is naturally expected for the $f_0(980)$ state
regardless of further details of particular models.

%%%%%%%%%%%%%%%%%%%%%%%%%%%%%%%%%%%%%%%%%%%%%%%%%%%%%%%%%%%%%%%%%%%%%%%%%%%
\section{Conclusion}  \label{Concl}
%%%%%%%%%%%%%%%%%%%%%%%%%%%%%%%%%%%%%%%%%%%%%%%%%%%%%%%%%%%%%%%%%%%%%%%%%%%

  The decay $\phi\to\gamma\pi\pi$ has been studied in an exactly solvable
coupled channel model containing the $\pi\pi$, $K\bar{K}$, and $q\bar{q}$
channels using separable potentials.  The $f_0(980)$ resonance corresponds
to {\it one} $S$-matrix pole close to the $K\bar{K}$ threshold; this pole has 
a dynamical origin and represents the molecular-like $K\bar{K}$ state.
  The molecular picture of the $f_0(980)$ meson is found to be in a fair
agreement with the experimental data.
We confirm the assessment of \cite{CIK93} that the earlier
conclusions about suppression of the $\phi\to\gamma f_0(980)$
branching ratio in the molecular $K\bar{K}$ model were partly
related to differences in modeling.

  The lightest scalar meson, $\sigma$, has a dynamical origin resulting from
the attractive character of the effective $\pi\pi$ interaction, with a partial
contribution from the coupling via the intermediate
scalar $q\bar{q}$ states.  The distinction between genuine $q\bar{q}$ states
and dynamical resonances, $\sigma$ and $f_0(980)$, can be illuminated by
considering the limit $N_c\to\infty$
where the $q\bar{q}$ states turn into infinitely narrow resonances while the
dynamical states disappear altogether.

  The structure of the $q\bar{q}$ state embedded into the mesonic continuum
has been analyzed using the calculated $q\bar{q}$ spectral density.
The gross structure of the quark--antiquark spectral density $\rho(s)$ is related
to the $f_0(1370)$ resonance.
  There is also a significant contribution to $\rho(s)$ in the low mass region
($\sigma$ meson) which is related to the strong coupling between
the $\pi\pi$ and $q\bar{q}$ channels.
  The same approach can also be used for the QCD sum rules
related to the gluon condensate by extending the coupled channel model to
include the mixing with the scalar glueballs.  The consideration of this topic
is beyond the scope of this paper.

%%%%%%%%%%%%%%%%%%%%%%%%%%%%%%%%%%%%%%%%%%%%%%%%%%%%%%%%%%%%%%%%%%%%%%%%%%%
\section*{Acknowledgments}
%%%%%%%%%%%%%%%%%%%%%%%%%%%%%%%%%%%%%%%%%%%%%%%%%%%%%%%%%%%%%%%%%%%%%%%%%%%

The author thanks M.P.~Locher for a fruitful collaboration which lead to
this paper, S.I.~Eidelman for bringing attention to the problem of the 
radiative $\phi$ decays and a discussion of the experimental data,
D.~Bugg and B.S.~Zou for a discussion of the nature of the $f_0(980)$.

%%%%%%%%%%%%%%%%%%%%%%%%%%%%%%%%%%%%%%%%%%%%%%%%%%%%%%%%%%%%%%%%%%%%%%%%%%%
%\renewcommand{\appendix}{%
%   \renewcommand{\section}{\secdef\Appendix\sAppendix}%
%   \setcounter{section}{0}%
%   \renewcommand{\thesection}{\Alph{section}}%
%}
%
%%%%%%%%%%%%%%%%%%%
\appendix
%%%%%%%%%%%%%%%%%%%
\setcounter{section}{0}%
\renewcommand{\thesection}{Appendix \Alph{section}}%

%%%%%%%%%%%%%%%%%%%%%%%%%%%%%%%%%%%%%%%%%%%%%%%%%%%%%%%%%%%%%%%%%%%%%%%%%%%
\section{The function $I(a,b)$ and the decay widths}
\label{AppIab}
%%%%%%%%%%%%%%%%%%%%%%%%%%%%%%%%%%%%%%%%%%%%%%%%%%%%%%%%%%%%%%%%%%%%%%%%%%%

The function $I(a,b)$ is given by (see e.g. \cite{LP90} and references
therein)
\begin{eqnarray}
  I(a,b) & = &
  \frac{1}{2(a-b)} - \frac{2}{(a-b)^2}
                     \left( f(\frac{1}{b})-f(\frac{1}{a}) \right)
                   + \frac{a}{(a-b)^2}
                     \left( g(\frac{1}{b})-g(\frac{1}{a}) \right)
  \label{Iab}
\\
  f(x) & = & \left\{
  \begin{array}{ll}
    -(\arcsin{\frac{\displaystyle 1}{\displaystyle 2\sqrt{x}}})^2
        \quad, & x>\frac{1}{4} \\
    \frac{1}{4} \left(
    \ln{\frac{\displaystyle 1+\sqrt{1-4x^2}}{\displaystyle 1-\sqrt{1-4x^2}}}
    - i\pi \right)^2
        \quad, & x \leq \frac{1}{4}
  \end{array} \right.
  \label{fx}
\\
  g(x) & = & \left\{
  \begin{array}{ll}
    \sqrt{4x^2-1} \arcsin{\frac{\displaystyle 1}{\displaystyle 2\sqrt{x}}}
        \quad, & x>\frac{1}{4} \\
    \frac{1}{2} \sqrt{1-4x^2} \left(
    \ln{\frac{\displaystyle 1+\sqrt{1-4x^2}}{\displaystyle 1-\sqrt{1-4x^2}}}
    - i\pi \right)
        \quad, & x \leq \frac{1}{4}
  \end{array} \right.
  \label{gx}
\end{eqnarray}

  The $\phi K^+K^-$ coupling constant $g_{\phi}$ is related to the decay
width by
\begin{eqnarray}
  \Gamma(\phi\to K^+K^-) & = &
  \frac{g_{\phi}^2}{48 \pi m_{\phi}^2}(m_{\phi}^2 - 4 m_K^2)^{3/2} \quad .
\label{GphiKK}
\end{eqnarray}
The relation between the coupling constants and decay widths for the
scalar meson $f_0$ has the form
\begin{eqnarray}
  \Gamma(f_0\to \pi\pi) & = & \frac{1}{2} \cdot
  \frac{g_{f_0\pi\pi}^2}{16 \pi m_{f_0}^2}(m_{f_0}^2 - 4 m_{\pi}^2)^{1/2}
\label{Gf0pipi}
\end{eqnarray}
where the extra factor $\frac{1}{2}$ accounts for the identity of the
two pions in the final state.

%\begin{eqnarray}
%   \Gamma_{f_0} & = & \Gamma_{f_0\to\pi\pi} =
%        \frac{g_{f_0\pi\pi}^2 k_{\pi\pi}(m_{f_0})}{8\pi m_{f_0}^2}
%\end{eqnarray}

%%%%%%%%%%%%%%%%%%%%%%%%%%%%%%%%%%%%%%%%%%%%%%%%%%%%%%%%%%%%%%%%%%%%%%%%%%%
\section{The Coupled Channel Model} \label{AppCCM}
%%%%%%%%%%%%%%%%%%%%%%%%%%%%%%%%%%%%%%%%%%%%%%%%%%%%%%%%%%%%%%%%%%%%%%%%%%%

% Green's function

The free Green function $G^0(s)$ is a diagonal matrix:
\begin{eqnarray}
   \mbox{\boldmath$G$}^0(s) & = &
   \left(
     \matrix{ \mbox{\boldmath$G$}^0_{1}(s) & 0 & 0 \cr
        0 & \mbox{\boldmath$G$}^0_{2}(s) & 0 \cr
        0 & 0 & \mbox{\boldmath$G$}^0_{3}(s) \cr }
   \right)
\label{G0}
\end{eqnarray}
where the single--channel Green functions have the form
\begin{eqnarray}
  \mbox{\boldmath$G$}^0_{1}(s) & = & \frac{2}{\pi}\;
  \int_{0}^{\infty} \frac{|k_1 \rangle\langle k_1|}{s/4-(m_\pi^2+k_1^2)}
                    k_1^2 dk_1 \\
  \mbox{\boldmath$G$}^0_{2}(s) & = & \frac{2}{\pi}\;
  \int_{0}^{\infty} \frac{|k_2 \rangle\langle k_2| }{s/4-(m_K^2+k_2^2)}
                    k_2^2 dk_2 \\
  \mbox{\boldmath$G$}^0_{3}(s) & = &
  %  \frac{|q\bar{q} \rangle\langle q\bar{q}| }{s-M_r^2}
  G^0_3(s) |q\bar{q} \rangle\langle q\bar{q}|
  \quad , \quad
   G^0_3(s) = \frac{1}{s-M_r^2}   \quad .
\label{GGG}
\end{eqnarray}
Here $|k_1\rangle$ and $|k_2\rangle$ denote the free $\pi\pi$ and $K\bar{K}$
states with relative momenta $k_1$ and $k_2$, respectively.
The state $|q\bar{q}\rangle$ in channel 3 has a bare mass $M_r$.
  With the form factors given by Eq.(\ref{FF}) the matrix elements of the
Green functions are
\begin{eqnarray}
  G^0_n(s) =
  \langle n | \mbox{\boldmath$G$}_n^0(s) | n \rangle & = &
  \frac{\lambda_n^3}{2(k_n(s) + i \lambda_n)^2} \ \ , \ \ n=1,2
\label{GME}
%\\
%  G^0_3(s) & = & \frac{1}{s-M_r^2}
%\label{GM3}
\end{eqnarray}
where $k_n(s)$ is the relative momentum in the channel $n$:
\begin{eqnarray}
   k_1(s) & = & \sqrt{s/4 - m_{\pi}^2} \label{k1s} \\
   k_2(s) & = & \sqrt{s/4 - m_K^2}     \label{k2s} \ \ .
\end{eqnarray}

The $\pi\pi$ elastic scattering amplitude $f_{\pi\pi}(s)$
and the $\pi\pi-K\bar{K}$ amplitude $f_{\pi\pi-K\bar{K}}(s)$
have the form:
\begin{eqnarray}
   f_{\pi\pi}(s) & = &
     - \langle k_1 | T(s) | k_1 \rangle =
     - \xi(k_1)^2 \;\frac{N_{11}(s)}{D(s)}
\label{f11}
\\
  f_{\pi\pi-K\bar{K}}(s) & = &
     - \langle k_1 | T(s) | k_1 \rangle =
     - \xi(k_1)\xi(k_2) \;\frac{N_{12}(s)}{D(s)}
\label{f12}
\end{eqnarray}
where
\begin{eqnarray}
   D(s) & = & 1 - u_{11}(s) G^0_1(s) - u_{22}(s) G^0_2(s) \nonumber\\
        &   &   + ( u_{11}(s) u_{22}(s) - u_{12}^2(s) ) G^0_1(s) G^0_2(s)
\label{D}
\\
   N_{11}(s) & = & u_{11}(s) - ( u_{11}(s) u_{22}(s) - u_{12}^2(s) ) G^0_3(s)
\label{N11}
\\
   N_{12}(s) & = & u_{12}  \label{N12}
\\
   u_{11}(s) & = & v_{11}(s) + \frac{g_{13}^2}{s-M_r^2}      \label{U11}
\\
   u_{12}(s) & = & v_{12}(s) + \frac{g_{13} g_{23}}{s-M_r^2} \label{U12}
\\
   u_{22}(s) & = & v_{22}(s) + \frac{g_{23}^2}{s-M_r^2}  \quad .  \label{U22}
\end{eqnarray}

The connection between the partial wave $S$-matrix and the scattering
amplitude $f_{\pi\pi}$ is given by
\begin{eqnarray}
     S_{J=0}^{I=0}(s) = \eta_0^0(s) e^{2i\delta_0^0(s)}
          = 1 + 2 i k_1 f_{\pi\pi}(s)
\label{SM}
\end{eqnarray}
where $\delta_0^0(s)$ is the scattering phase and $\eta_0^0(s)$ is
the inelasticity parameter.
The $T$-matrix element is related to the amplitude (\ref{f12}) by
\begin{eqnarray}
     t_{K\bar{K}-\pi\pi} & = & 8\pi\sqrt{s} f_{K\bar{K}-\pi\pi} \quad .
\label{T12}
\end{eqnarray}

%%%%%%%%%%%%%%%%%%%%%%%%%%%%%%%%%%%%%%%%%%%%%%%%%%%%%%%%%%%%%%%%%%%%%%%%%%%
%\section{The Spectral Density of the $q\bar{q}$ State Embedded into Continuum}
%\label{AppSR}
%%%%%%%%%%%%%%%%%%%%%%%%%%%%%%%%%%%%%%%%%%%%%%%%%%%%%%%%%%%%%%%%%%%%%%%%%%%

The spectral density of the $q\bar{q}$ state has the form \cite{LMZ98}
\begin{eqnarray}
   \rho(s)   & = & \frac{1}{2\pi i}(G_3(s-i\epsilon)-G_3(s+i\epsilon)) \ \ ,
   \label{w}
\end{eqnarray}
where $G_3(s)$ is the exact Green function in the $|q\bar{q}\rangle$
subspace:
\begin{eqnarray}
   G_3(s) & = & \langle q\bar{q}|G(s)|q\bar{q} \rangle
                      = \frac{1}{s - M_r^2 - \Pi(s)}     \label{G3}
\end{eqnarray}
and $\Pi(s)$ is the mass operator of the $q\bar{q}$ state:
\begin{equation}
   \Pi(s)  = 
   \frac{4
   (g_{13}^2 G_1^0(s) + g_{23}^2 G_2^0(s) +
   (2v_{12}(s) g_{13}g_{23} - g_{13}^2 v_{22}(s) - g_{23}^2 v_{11}(s))
   G_1^0(s)G_2^0(s))}
   {1 - v_{11}(s) G_1^0(s) - v_{22}(s) G_2^0(s) +
       (v_{11}(s) v_{22}(s) - v_{12}^2(s)) G_1^0(s) G_2^0(s) }
   \ \ .
\label{MOG3}
\end{equation}
The spectral density $\rho(s)$ satisfies the normalization condition
\begin{eqnarray}
   \int_{4 m_{\pi}^2}^{\infty} \rho(s) ds
    & = &  1  \quad .
\label{SumR}
\end{eqnarray}
Equations (\ref{w},\ref{SumR}) represent the completeness relation
projected onto the $q\bar{q}$ channel and the normalization
$\langle q\bar{q}|q\bar{q}\rangle=1$.

%%%%%%%%%%%%%%%%%%%%%%%%%%%%%%%%%%%%%%%%%%%%%%%%%%%%%%%%%%%%%%%%%%%%%%%%%%%
\section{The Nonrelativistic Approximation} \label{AppNR}
%%%%%%%%%%%%%%%%%%%%%%%%%%%%%%%%%%%%%%%%%%%%%%%%%%%%%%%%%%%%%%%%%%%%%%%%%%%

The amplitudes corresponding to the diagrams in Fig.~\ref{FigMphigpipi}
have the form:
\begin{eqnarray}
  M^{NR} & = & \frac{e g_{\phi} \; t_{K^+K^--\pi\pi}}{(2\pi)^3}
          (I_a + I_b + I_c) =
          \frac{e g_{\phi} \; t_{K^+K^--\pi\pi} \; m_{\phi}\omega}{2\pi^2 m_K^2}
          \; J_{\lambda}(M_{\pi\pi})
\end{eqnarray}
where
\begin{eqnarray}
  I_a & = & 8
            \int \frac{d^3\bk}{2E_k}
            \frac{ (\beps_{\phi}\cdot\bk) (\beps_{\gamma}\cdot(\bk+\bq/2) )
                   F(|\bk+\bq/2|)
                 }{ (k^2-m_K^2+i\epsilon) ((k-q)^2-m_K^2+i\epsilon)} =
\\
    & = &  \frac{4 m_K}{ m_{\phi} M_{\pi\pi} }
           \int d^3\bk
           \frac{ (\beps_{\phi}\cdot\bk) (\beps_{\gamma}\cdot\bk)
                  F(|\bk+\bq/2|) }
                { (\Delta_{\phi} - \bk^2+i\epsilon)
                  (\Delta_{\pi\pi} - (\bk+\bq/2)^2+i\epsilon) }
\label{NRMa}
\\
 I_b & = &  2 (\beps_{\phi}\cdot\beps_{\gamma})
            \int \frac{d^3\bk}{2E_k}
            \frac{ F(|\bk+\bq/2|) }{ ((k-q)^2-m_K^2+i\epsilon) } =
\\
     & = &  \frac{ (\beps_{\phi}\cdot\beps_{\gamma}) }{ M_{\pi\pi} }
            \int d^3\bk
            \frac{F(|\bk|)}{ (\Delta_{\pi\pi} - \bk^2 +i\epsilon) }
\label{NRMb}
\\
 I_c & = &  2
            \int \frac{d^3\bk}{2E_k}
            \frac{ (\beps_{\phi}\cdot\bk) (\beps_{\gamma} \cdot \bk/k)
                   \frac{d}{d|\bk|} F(|\bk|) }
                 { (k^2 - m_K^2 +i\epsilon) } =
\\
     & = &  I_{c1} + I_{c2}
\label{NRMc}
\\
     I_{c1} & = &
           -\frac{ (\beps_{\phi}\cdot\beps_{\gamma}) }{ M_{\phi} }
            \int d^3\bk
            \frac{F(|\bk|)}{ (\Delta_{\phi} - \bk^2 +i\epsilon) }
\label{NRMc1}
\\
     I_{c2} & = &
           -\frac{ 2 (\beps_{\phi}\cdot\beps_{\gamma}) }{ 3 M_{\phi}}
            \int d^3\bk
            \frac{\bk^2 F(|\bk|)}{ (\Delta_{\phi} - \bk^2 +i\epsilon)^2}
\label{NRMc2}
\end{eqnarray}
Here $E_k = \sqrt{k^2+m_K^2}$,
     $\Delta_{\phi} = (m_{\phi} - 2m_K) m_K$,
     $\Delta_{\pi\pi} = (M_{\pi\pi} - 2m_K) m_K$,
the form factor $F(|\bk|)=\xi_2(|\bk|)$ according to Eq.(\ref{FF}),
the subscript $\lambda$ in $J_{\lambda}(M_{\pi\pi})$ refers to the
form--factor parameter.
In deriving the nonrelativistic approximation (\ref{NRMa}-\ref{NRMc})
we keep only the positive-energy parts of the kaon propagators and
substitute $d^3k/E_k\to d^3k/m_K$ in the final integral. The
total amplitude $M^{NR}$ vanishes at $\omega=0$ as expected
for the electric dipole transition. The diagram
Fig.\ref{FigMphigpipi}(c) contains two terms according to Eq.(\ref{NRMc})
which can be combined with the loop radiation term (\ref{NRMa})
and the contact term (\ref{NRMb}) in a way that the combinations
$(I_a+I_{c2})$ and $(I_b+I_{c1})$ are
explicitly finite and both vanish at $\omega=|\bq|\to 0$.
The integrals (\ref{NRMa}-\ref{NRMc2}) with our choice of the
form factor can be straightforwardly calculated in analytical form.

%%%%%%%%%%%%%%%%%%%%%%%%%%%%%%%%%%%%%%%%%%%%%%%%%%%%%%%%%%%%%%%%%%%%%%%%%%%

%%%%%%%%%%%%%%%%%%%%%%%%%%%%%%%%%%%%%%%%%%%%%%%%%%%%%%%%%%%%%%%%%%%%%%%%%%%

\end{document}